\documentclass[aps,prx,reprint,twocolumn,floatfix,nofootinbib]{revtex4-2}
\usepackage{epsfig,amsmath,amssymb,color,comment,physics, dsfont, bbold}
\usepackage[makeroom]{cancel}
\usepackage[caption=false]{subfig}
\usepackage{mathrsfs}
\usepackage[countmax]{subfloat}
\usepackage[normalem]{ulem}
\usepackage[english]{babel}
\usepackage{dsfont}
\usepackage{color}
\usepackage{float}
\usepackage[dvipsnames]{xcolor}
\usepackage{amsmath}
\usepackage{braket}
\usepackage{tikz}
\usetikzlibrary{quantikz2}

\usepackage[bookmarks=true,colorlinks,linkcolor=blue,urlcolor=NavyBlue,citecolor=RoyalBlue]{hyperref}

\newcommand\at[2]{\left.#1\right|_{#2}}
\newcommand{\ibmboston}{\texttt{ibm\_boston}}

\graphicspath{{./Paper_figures/}}
\begin{document}

\title{Simulating Black Hole Thermality and Interior Scrambling on a Superconducting Quantum Processor}

\author{Ryan Smith$^{*,\dagger}$}
\affiliation{School of Physics and Astronomy, University of Leeds, Leeds LS2 9JT, UK}

\author{Ewan Forbes$^{*}$}
\affiliation{School of Physics and Astronomy, University of Leeds, Leeds LS2 9JT, UK}

\author{Iason A. Sofos}
\affiliation{School of Physics and Astronomy, University of Leeds, Leeds LS2 9JT, UK}

\author{Andrew Hallam}
\affiliation{School of Physics and Astronomy, University of Leeds, Leeds LS2 9JT, UK}

\author{Jiannis K. Pachos$^{\ddagger}$}
\affiliation{School of Physics and Astronomy, University of Leeds, Leeds LS2 9JT, UK}

\date{\today}
\begin{abstract}

We implement a chiral spin-chain black hole simulator on IBM superconducting quantum hardware and probe, within a common microscopic framework, both semiclassical horizon physics and interacting quantum scrambling. We first measure the dispersion relation across the exterior, horizon and over-tilted interior regimes, reproducing the predicted evolution of the effective light-cone structure. To probe Hawking thermality, we prepare a localised excitation inside the horizon and monitor its density response at an exterior site, observing the predicted inverse relation between the peak arrival time and the surface gravity, thereby establishing a calibrated dynamical estimator of the Hawking temperature. Beyond the semiclassical regime, we continuously tune the interactions and distinguish non-exponential operator spreading in the free-fermion limit from Lyapunov-like OTOC decay in the strongly interacting chiral regime. These measurements use observable-specific Floquet circuits derived from the same parent chiral model, including its mean-field and coordinate-equivalent XY descriptions, to reduce circuit depth while preserving the physics relevant to each probe. Our results provide a unified programmable platform for studying horizon geometry, Hawking thermality and interacting scrambling on quantum hardware.

\end{abstract}
\maketitle

\begingroup
\renewcommand{\thefootnote}{\fnsymbol{footnote}}
\footnotetext[1]{These authors contributed equally to this work.}
\footnotetext[2]{
\href{mailto:ll17rps@leeds.ac.uk}{ll17rps@leeds.ac.uk}}
\footnotetext[3]{
\href{mailto:J.K.Pachos@leeds.ac.uk}{j.k.pachos@leeds.ac.uk}}
\endgroup

\section{Introduction}

Analogue-gravity experiments have been proposed~\cite{PhysRevLett.46.1351,Fedichev2003GibbonsHawking, Fedichev2004Observerdependence, Barcelo_et_al_2005, Philbin_et_al_2008, Gooding_et_al_2020, Tian2022trappedionshawking, benhemou2025opticalblackhole} and realised~\cite{Weinfurtner_et_al_2011,Steinhauer2016, de_Nova_et_al_2019, Eckel_et_al_2018, Wittemer_et_al_2019, de_Nova_et_al_2019, Hu_et_al_2019, Viermann_et_al_2022, Jacquet_et_al_2022, Shi_et_al_2023} to demonstrate Hawking-like emission and its thermal character in controlled laboratory settings. However, accessing both semiclassical horizon physics and strongly interacting scrambling within the same programmable platform remains challenging. Recent advances in quantum processors enable many-body dynamics to be implemented and measured at the circuit level, including experimental probes of information scrambling using out-of-time-ordered correlators (OTOCs)~\cite{Landsman2019Verifiedscrambling, Joshi2020TrappedIonscrambling, Blok2021Qutritscrambling, Green2022FinitetempOTOC, Google2024OTOC, Google2025OTOC, Seki2025FloquetScrambling}. Quantum hardware, therefore, offers a promising route for simulating black hole-inspired dynamics beyond the regimes accessible to analytical or classical numerical methods.

Several complementary approaches have advanced the quantum simulation of black hole and scrambling physics. Superconducting-qubit platforms have been used to simulate Hawking radiation in an on-chip black hole analogue~\cite{Shi_et_al_2023}, while programmable quantum processors have enabled the study of light-cone propagation in emergent curved spacetimes~\cite{rhyno2026curvedspacetimedynamics}. In parallel, superconducting processors have established OTOCs and echo protocols as practical probes of scrambling, operator growth, and quantum chaos~\cite{Google2024OTOC,Google2025OTOC,algorithmiq2026TEM}. These works demonstrate the relevant ingredients separately. Here, we combine them within a single black hole spin-chain framework comprising a semiclassical sector for Hawking thermality and an interacting chiral sector for interior scrambling.

Black holes provide a unique setting in which quantum physics, thermodynamics, and gravity meet. The prediction that black holes emit thermal radiation at a temperature fixed by their surface gravity, and the apparent information loss associated with this process, have led to a longstanding tension between black hole evaporation and unitary quantum evolution~\cite{Bekenstein1973,Hawking1975,Hawking_1976,Hawking_1976Breakdown,Page1993,AMPS2013,wald_hawking,Page_2005}. At the same time, black holes are expected to be exceptionally efficient scramblers of quantum information, with their chaotic dynamics diagnosed by OTOCs and constrained by the universal chaos bound~\cite{SekinoSusskind2008,ShenkerStanford2014,MaldacenaShenkerStanford2016}. These connections motivate the study of Hawking thermality and many-body scrambling within a common framework, particularly in relation to information-recovery scenarios such as the Page curve and the Hayden-Preskill protocol~\cite{Page1993,HaydenPreskill2007}.

\begin{figure*}
    \centering
    \includegraphics[width=0.99\linewidth]{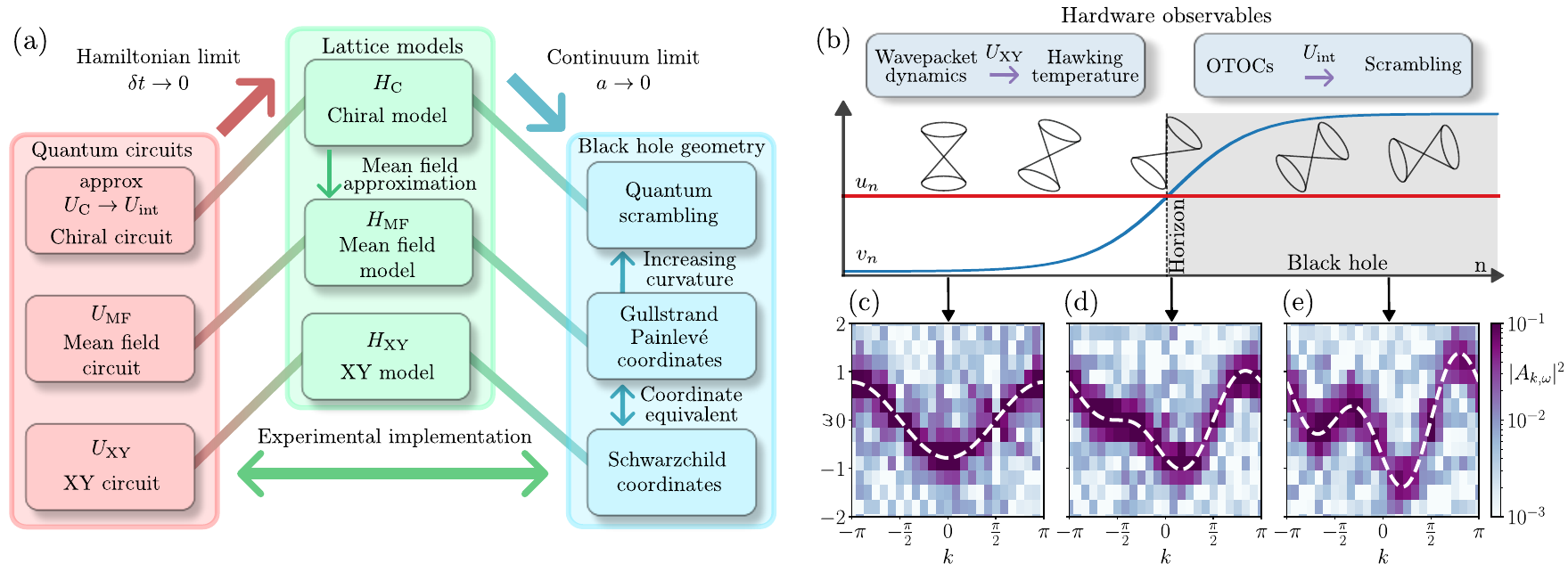}
\caption{
Overview of the black hole quantum simulator and its hardware validation.
(a) Relations among the lattice Hamiltonians, quantum circuits and effective theories considered in this work. The interacting chiral model $H_{\rm C}$ reduces under mean-field theory to $H_{\rm MF}$, whose semiclassical geometry is coordinate-equivalent to that of the free-fermion XY model $H_{\rm XY}$. The circuits $U_{\rm MF}$ and $U_{\rm XY}$ reproduce their respective Hamiltonian dynamics in the small-time-step limit. The reduced circuit $U_{\rm int}$ instead isolates the non-free-fermionic sector of the parent chiral circuit used to probe scrambling.
(b) Spatial coupling profile and effective light-cone structure across the exterior, horizon and interior. The lower-depth XY circuit is used to extract the Hawking temperature from the horizon and exterior regions, while the interacting chiral circuit is used to probe scrambling through OTOCs in the interior region.
(c)–(e) Dispersion relations of $U_{\rm MF}$ measured on superconducting quantum hardware for $\theta^{(u)}=\pi/4$ and, respectively, $\theta^{(v)}=0$, $\pi/8$ and $\pi/4$. These regimes correspond to the exterior, horizon and interior, where the dispersion is untilted, critical and over-tilted. The white dashed curves show the analytical prediction in Eq.~\eqref{eq:dispersion1}. The quasi-energy scale is set by the Floquet time step $\delta t$. All measurements were performed for $N=24$ sites on \ibmboston{}, using $n_s=10000$ shots together with Pauli twirling, measurement readout twirling and zero-noise extrapolation.}
    \label{fig:overview}
\end{figure*}

We realise this programme using a spin-chain black hole analogue implemented on IBM superconducting quantum hardware. The regimes considered in this work are summarised in Fig.~\ref{fig:overview}(a). Our starting point is the interacting chiral spin-chain $H_{\rm C}$, whose semiclassical mean-field limit $H_{\rm MF}$ describes Dirac fermions propagating in an effective curved spacetime with an event horizon, while its fully interacting regime exhibits fast scrambling~\cite{HornerHallamPachos2023, Forbes2023interactingchiralmodel, Daniel2025chiralscrambling, daniel2025blackholeteleport}. We implement Trotterised quantum circuits for three related lattice realisations: the interacting chiral circuit $U_{\rm C}$, the mean-field chiral circuit $U_{\rm MF}$, and a free-fermion XY circuit $U_{\rm XY}$. The effective black hole geometry can be simulated using either the chiral model, where it is realised in Gullstrand-Painlevé coordinates, or the XY model, where it is realised in Schwarzschild coordinates and reduces to the Rindler metric near the horizon. When their coupling profiles are matched, the two descriptions realise equivalent semiclassical geometries and therefore predict the same Hawking temperature.

In principle, the complete interacting chiral circuit $U_{\rm C}$ provides a unified microscopic simulator of the exterior, horizon, and interior regimes. Implementing this parent circuit would allow semiclassical geometry, Hawking thermality, and interacting scrambling to be studied without changing the underlying model. It would also provide access to system sizes for which generic interacting real-time dynamics becomes increasingly costly to simulate classically. On present hardware, however, the depth of the complete circuit would obscure the relevant signals. We therefore use physically motivated reductions of $U_{\rm C}$, tailored to the observable being measured.

This observable-tailored strategy is particularly useful on current quantum hardware. The mean-field circuit $U_{\rm MF}$ retains the single-particle dynamics needed to verify the effective light-cone structure. The coordinate-equivalent XY circuit $U_{\rm XY}$ has a substantially lower circuit depth and is therefore better suited to extracting the Hawking temperature with reduced hardware error. By contrast, the reduced interacting circuit $U_{\rm int}$ retains the many-body terms needed to probe scrambling. We consequently use $U_{\rm MF}$ to measure the dispersion relation, $U_{\rm XY}$ to probe Hawking thermality, and $U_{\rm int}$ to measure infinite-temperature OTOCs and their Lyapunov growth.

The remainder of this paper is organised as follows: In Sec.~\ref{sec:chiral}, we introduce the spin-chain models and Trotterised circuits used in this work and explain how their continuum and mean-field limits realise effective black hole geometries. In Sec.~\ref{sec:Dispersion relation}, we present IBM hardware results for the Floquet mean-field chiral circuit $U_{\rm MF}$ and show that it reproduces the dispersion relation outside, at, and inside the black hole horizon. In Sec.~\ref{sec:Hawking temperature}, we use the lower-depth XY circuit $U_{\rm XY}$ to extract the Hawking temperature from the dynamics of a wave packet near the horizon. Finally, in Sec.~\ref{sec:scrambling}, we use the experimentally implementable reduction $U_{\rm int}$ of $U_{\rm C}$ to continuously tune the dynamics between the free-fermion and interacting chiral limits. By measuring infinite-temperature OTOCs across this interpolation, we resolve the onset of interaction-driven scrambling and extract the Lyapunov exponent in the strongly interacting regime.

\section{Black hole spin-chain simulators and circuit implementation}
\label{sec:chiral}

We simulated several central properties of black holes within a common spin-chain framework. Experimentally, we implement brick-wall quantum circuits on \ibmboston{}. The circuit unitaries are defined at a finite time step $\delta t$ and approach Hamiltonian time evolution as $\delta t\rightarrow0$. Section~\ref{subsec:spinchain} introduces the chiral spin-chain, whose mean-field continuum limit describes freely propagating fermions in a Schwarzschild black hole spacetime expressed in Gullstrand-Painlevé coordinates. The continuum limit of the XY spin-chain describes the same semiclassical geometry in Schwarzschild coordinates. We introduce the corresponding Floquet model in Sec.~\ref{subsec:floquet_chiral} and the observable-specific circuit implementations in Sec.~\ref{subsec: chiral circuit encoding}.

\subsection{The chiral lattice model simulator}
\label{subsec:spinchain}

Lattice regularisations of curved spacetime quantum field theories have long been used to investigate the robustness of Hawking emission~\cite{CorleyJacobson1998} and, more recently, position-dependent hopping models have realised synthetic horizons, horizon-induced thermality and Hawking spectra in both Hamiltonian and Floquet dynamics~\cite{morice2022quantum, Mertens2022, Maertens_2024}.

As our starting point, we consider the spin-$1/2$ chiral spin-chain Hamiltonian
\begin{equation}\label{eq:chiralHamiltonian}
H_{\rm C} = \frac{1}{a}\sum_{n=1}^{N} \left[ -\frac{u_n}{2}
\left(X_nX_{n+1} + Y_nY_{n+1} \right) + \frac{v_n}{4}\chi_n \right],
\end{equation}
where $u_n$ and $v_n$ are real-valued, spatially varying couplings and
\begin{equation}
\chi_n=\boldsymbol{\sigma}_n\cdot
\left(\boldsymbol{\sigma}_{n+1}\times\boldsymbol{\sigma}_{n+2}\right)
\label{Equation: Chiral Operator}
\end{equation}
is the SU(2)-invariant scalar spin-chirality operator, which measures the handedness of three-spin correlations~\cite{jiannis_optical_chiral, Tsomokos2008chiral}. Here, $\boldsymbol{\sigma}_n =( X_n, Y_n, Z_n)$ with $\{X_n,Y_n,Z_n\}$ denoting the Pauli matrices at site $n$, $a=L/N$ is the lattice spacing, and $x=na$ becomes the continuum position as $a\rightarrow0$ at fixed $L$. 

As demonstrated in Appendix~\ref{Appendix:JW mean field}, applying a mean-field decoupling to Eq.~\eqref{eq:chiralHamiltonian} results in the following mean-field Hamiltonian:
\begin{equation}
\label{eq:MFHamiltonian}
\begin{aligned}
H_{\rm MF} = -\frac{1}{a}&\sum_{n=1}^{N} \biggl[\frac{u_n}{2}
\left(X_nX_{n+1} + Y_nY_{n+1} \right) \\
&+ \frac{v_n}{4}\left(X_nZ_{n+1}Y_{n+2}- Y_nZ_{n+1}X_{n+2}\right) \biggr],
\end{aligned}
\end{equation}
for which the dispersion describing the fermionic excitations in this system is
\begin{equation}
\label{eq:dispersion1}
\omega(k) = v\sin(2k) \pm 2u\cos(k).
\end{equation}
The two terms in the dispersion of Eq.~\eqref{eq:dispersion1} have distinct geometric roles. The XY coupling $u$ sets the symmetric hopping and, hence, the local propagation speed, while the chiral coupling $v$ tilts the dispersion and, therefore, the Dirac cones that emerge about the Fermi-points \cite{HornerHallamPachos2023, Forbes2023interactingchiralmodel}.

It has been demonstrated that the mean-field Hamiltonian's low-energy continuum limit is described by massless Dirac fermions propagating freely in an effective curved spacetime~\cite{HornerHallamPachos2023, Forbes2023interactingchiralmodel, sofos2025chiralentanglement}. For slowly varying couplings, the line element of this spacetime is that of the Gullstrand-Painlevé metric
\begin{equation}
ds^2_{\rm GP}=\left(1-\frac{v(x)^2}{u(x)^2}\right)dt^2-\frac{2v(x)}{u(x)^2}dt\,dx-\frac{dx^2}{u(x)^2},
\label{eq:GPmetric}
\end{equation}
which has two event horizons at $v(x_{h})=\pm u(x_{h})$. The regions $|v|<|u|$ and $|v|>|u|$ correspond, respectively, to the exterior and interior. In the interior region, the effective Dirac cone of Eq.~\eqref{eq:dispersion1} is over-tilted. By introducing the Schwarzschild time coordinate $\tilde{t}$ through $d\tilde t=dt-\frac{v(x)}{u(x)^2-v(x)^2}\,dx$, the Gullstrand-Painlev\'e metric can be diagonalised to give
\begin{equation}
\label{eq:DiagonalMetricGeneral}
ds^2=\frac{1}{u(x)} \left(f(x)d\tilde t^{\,2}-\frac{dx^2}{f(x)}\right),
\quad
f(x)=u(x)-\frac{v(x)^2}{u(x)}.
\end{equation}
which is conformally equivalent to the Schwarzschild metric and, thus, has the same surface gravity \cite{Jacobson_1993}. Unlike the Gullstrand-Painlevé metric, which is regular across the horizon, this diagonal metric has a coordinate singularity at $f(x_{h})=0$, and is therefore only valid outside the black hole. Despite this, both coordinate descriptions encode the same exterior spacetime, with scalar quantities being invariant between them. Additionally, near the horizon, the line element of Eq.~\eqref{eq:DiagonalMetricGeneral} reduces locally to that of the Rindler metric.

This coordinate equivalence motivates a lower-depth lattice realisation. Setting the chirality to zero and encoding $u_n\sim f(na)$ in a position-dependent nearest-neighbour coupling gives
\begin{equation}
\label{eq:XYHamiltonian}
H_{\rm XY}=-\frac{1}{a}\sum_{n=1}^{N}\frac{u_n}{2}\left(X_nX_{n+1}+Y_nY_{n+1}\right).
\end{equation}
Its low-energy continuum limit realises the same diagonal Schwarzschild geometry as the mean-field chiral model when the coupling profiles are appropriately matched. The two descriptions, therefore, have the same horizon position, surface gravity, and Hawking temperature. Because $H_{\rm XY}$ contains only nearest-neighbour hopping, it admits a substantially lower-depth circuit and is used below for the hardware measurement of the Hawking temperature. A related position-dependent hopping model has been shown to realise synthetic horizons and horizon-induced thermality~\cite{Mertens2022,morice2022quantum}

The full chiral Hamiltonian $H_{\rm C}$ retains the interacting many-body dynamics absent from $H_{\rm MF}$ and $H_{\rm XY}$, thereby extending the black hole analogue beyond the non-interacting semiclassical description and enabling the study of scrambling through OTOCs. Previous work has demonstrated that the mean-field theory accurately captures the weakly interacting exterior regime, $|v(x)|<|u(x)|$, whereas interactions become important in the interior, $|v(x)|>|u(x)|$~\cite{HornerHallamPachos2023}. We therefore use the XY and mean-field circuits as semiclassical benchmarks for the horizon geometry and Hawking temperature, while the interacting chiral circuit probes the fast scrambling characteristic of the black hole interior. This division of roles is summarised in Fig.~\ref{fig:overview}(a) and (b). The exterior, horizon, and interior regimes can furthermore be distinguished through the dispersion relation in Eq.~\eqref{eq:dispersion1}, whose hardware measurement is shown in Fig.~\ref{fig:overview}(c)-(e) and discussed further in Sec.~\ref{sec:Dispersion relation}.

\subsection{The analogue Floquet chiral simulator}
\label{subsec:floquet_chiral}

To implement the dynamics of the chiral Hamiltonian on quantum hardware, we approximate the time-evolution operator generated by Eq.~\eqref{eq:chiralHamiltonian} using a first-order Trotter decomposition. Writing $t=m_{t}\delta t$, the evolution is expressed as
\begin{equation}
U(t)
=
\left[U(\delta t)\right]^{m_{t}},
\end{equation}
where a single Floquet step is
\begin{equation}
\label{eq:floquetchiralop}
\begin{aligned}
U(\delta t)
&=
\exp\Biggl(
- {i} \sum_{n=1}^{N} \frac{\theta^{(v)}_n}{2} \chi_{n}
\Biggr)
\\
&\quad \times
\exp\Biggl(
-i \sum_{n=1}^{N}
\frac{\theta^{(u)}_n}{2}\big(
X_{n}X_{n+1} + Y_{n}Y_{n+1}
\big)
\Biggr)
\\
&\quad + \mathcal{O}\left(\delta t^{2}\right),
\end{aligned}
\end{equation}
with $\theta^{(u)}_n=-u_n\delta t$ and $\theta^{(v)}_n=v_n\delta t/2$. We label the total time evolved as $T=M_t\delta t$. The relative factor of $1/2$ in these definitions maps the mean-field horizon condition $|v_n|=|u_n|$ to
\begin{equation}
|\theta^{(v)}| = \frac{|\theta^{(u)}|}{2}.
\end{equation}
Accordingly, $|\theta^{(v)}_n|>|\theta^{(u)}_n|/2$ and $|\theta^{(v)}_n|<|\theta^{(u)}_n|/2$ correspond to the interior and exterior regimes of the emergent black hole geometry, respectively, as illustrated in Fig.~\ref{fig:overview}(b).

Although Eq.~\eqref{eq:floquetchiralop} originates from a Trotterisation of the Hamiltonian dynamics, we subsequently treat $\theta^{(u)}_n$ and $\theta^{(v)}_n$ as independently tunable Floquet parameters. Their ratio controls the effective spacetime regime, while the Hamiltonian evolution is recovered in the small-angle limit. To obtain hardware-compatible circuits, the chiral operator's contribution to the time evolution is further decomposed into local three-body Pauli terms. The complete decomposition and the resulting parent Floquet unitary $U_{\rm C}$ are given in Appendix~\ref{Appendix: Circuit decompositions}.

\subsection{Circuit encoding of chiral unitary circuits}
\label{subsec: chiral circuit encoding}

Having defined the parent chiral unitary $U_{\rm C}$, we now introduce the circuit families used in the hardware experiments. In principle, $U_{\rm C}$ contains both the semiclassical horizon dynamics and the interacting many-body physics required to describe the exterior, horizon, and interior within a single microscopic model. A direct implementation of this complete evolution would therefore provide the most unified simulation of the black hole analogue. Moreover, quantum hardware offers a natural route to extending such interacting dynamics beyond the system sizes that are accessible to generic exact classical simulations.

On present noisy quantum processors, however, the depth of the complete chiral circuit limits the accessible evolution time and reduces the visibility of the measured signals~\cite{Preskill2018quantumcomputingin, RevModPhys.94.015004}. Therefore, we tailor the circuit to the observable of interest, using physically motivated reductions of the parent model rather than attempting to implement every term simultaneously. The mean-field circuit $U_{\rm MF}$ retains the single-particle chiral dynamics required to reconstruct the dispersion relation. For the Hawking-temperature measurement, we further exploit the coordinate equivalence of the semiclassical geometry and use the lower-depth XY circuit $U_{\rm XY}$. Finally, the interacting circuit $U_{\rm int}$ retains the non-free-fermionic sector responsible for the early-time scrambling dynamics. 

The basic building blocks for these circuits can be given in terms of the local Hamiltonians
\begin{equation}
\begin{split}
h^{(0)}_{n}=&~ X_n X_{n+1} + Y_n Y_{n+1}, \\
h^{(1)}_{n}=&~ Z_n X_{n+1} Y_{n+2}-Z_n Y_{n+1} X_{n+2},\\
h^{(2)}_{n}=&~ Y_n Z_{n+1} X_{n+2}-X_n Z_{n+1} Y_{n+2},\\
h^{(3)}_{n}=&~ X_n Y_{n+1} Z_{n+2}-Y_n X_{n+1} Z_{n+2},
\end{split}
\label{Equation: Local Hamiltonians}
\end{equation}
which allow the chiral operator to be expressed as
\begin{equation}
\chi_n=h^{(1)}_n+h^{(2)}_n+h^{(3)}_n.
\end{equation}
From the local Hamiltonians of Eq.~\eqref{Equation: Local Hamiltonians}, we can further define the corresponding local unitaries as
\begin{equation}
u^{(i)}_n(\theta)=\exp\left[-i\frac{\theta}{2}h^{(i)}_n\right],
\label{Equation: Local Trotterisation Unitaries}
\end{equation}
where we use the notation $h^{(i)}$ to denote the location of the $Z$ operator in the Pauli string. 

Considering first the semiclassical mean-field contributions to the dynamics of~\eqref{eq:floquetchiralop}, it is shown in Appendix~\ref{Appendix: Circuit decompositions} that the mean-field circuit is given by
\begin{equation} \label{eq:MFfloquetcircuit}
\begin{aligned}
&U_{\mathrm{MF}}
={}
\prod_{\scriptscriptstyle n\equiv2\,(\mathrm{mod}\,3)} u^{(2)}_n(\theta^{(v)}_n)
\prod_{\scriptscriptstyle n\equiv1\,(\mathrm{mod}\,3)} u^{(2)}_n(\theta^{(v)}_n)
\\
&\times \prod_{\scriptscriptstyle n\equiv0\,(\mathrm{mod}\,3)} u^{(2)}_n(\theta^{(v)}_n)
\prod_{\scriptscriptstyle n,\rm{odd}} u^{(0)}_n(\theta^{(u)}_n)
\prod_{\scriptscriptstyle n, \rm{even}} u^{(0)}_n(\theta^{(u)}_n),
\end{aligned}
\end{equation}
where mutually commuting terms have been arranged into brick-wall layers. The dynamics of $U_{\mathrm{MF}}$ approaches that of the mean-field Hamiltonian in Eq.~\eqref{eq:MFHamiltonian} as $\delta t\rightarrow0$. This circuit is used below to verify the effective light-cone structure through spectroscopy.

For the Hawking-temperature protocol, we use the lower-depth XY circuit
\begin{equation} \label{eq:XYfloquetcircuit}
\begin{aligned}
&U_{\mathrm{XY}}=
\prod_{\scriptscriptstyle n,\rm{odd}} u^{(0)}_n(\theta^{(u)}_n)
\prod_{\scriptscriptstyle n, \rm{even}} u^{(0)}_n(\theta^{(u)}_n),
\end{aligned}
\end{equation}
which is obtained from $U_{\mathrm{MF}}$ by setting $\theta^{(v)}_n=0$ and approaches the XY Hamiltonian in Eq.~\eqref{eq:XYHamiltonian} in the small-time-step limit.

To probe scrambling, we instead retain the non-free-fermionic terms of Eq.~\eqref{eq:floquetchiralop}. The resulting interacting circuit is
\begin{equation}
\begin{split}
& U_{\rm int} =\\
&
\prod_{\scriptscriptstyle n\equiv2\,(\mathrm{mod}\,3)} u^{(3)}_n(\theta^{(v)}_n)
\prod_{\scriptscriptstyle n\equiv1\,(\mathrm{mod}\,3)} u^{(3)}_n(\theta^{(v)}_n)
\prod_{\scriptscriptstyle n\equiv0\,(\mathrm{mod}\,3)} u^{(3)}_n(\theta^{(v)}_n)
\\
&\times
\prod_{\scriptscriptstyle n\equiv2\,(\mathrm{mod}\,3)} u^{(1)}_n(\theta^{(v)}_n)
\prod_{\scriptscriptstyle n\equiv1\,(\mathrm{mod}\,3)} u^{(1)}_n(\theta^{(v)}_n)
\prod_{\scriptscriptstyle n\equiv0\,(\mathrm{mod}\,3)} u^{(1)}_n(\theta^{(v)}_n). \label{eq:int_gate_structure}
\end{split} 
\end{equation}
This reduced circuit isolates the interacting sector responsible for the early-time scrambling behaviour, which is studied in Sec.~\ref{sec:scrambling}. 

The three circuits we implement should not be viewed as independent simulators, but as observable-specific reductions of a common parent chiral model, chosen to balance physical fidelity against the circuit depths accessible on current hardware. In the following subsection, we introduce a tunable generalisation of $U_{\rm int}$ that continuously interpolates between free-fermion and interacting chiral dynamics, allowing us to resolve the emergence of chaotic scrambling as interactions are switched on.

\subsection{Tuning between free-fermion and interacting chiral dynamics}

To resolve how interactions generate scrambling, we introduce a one-parameter generalisation of $U_{\text{int}}$ that continuously interpolates between free-fermion and interacting chiral dynamics. Specifically, we replace the controlled-$Z$ gates that conjugate the $XX+YY$ rotation with controlled-phase gates
$CP(\phi)=\text{diag}(1,1,1,e^{i\phi})$. Since $CP(0)$ is the identity and $CP(\pi)=CZ$, the phase $\phi$ controls the interaction-generating conditional phase while leaving the remaining circuit structure unchanged. Therefore, the local gates $u^{(1)}_n(\theta^{(v)}_n)$ and $u^{(3)}_n(\theta^{(v)}_n)$ take the modified forms
\begin{equation}
\label{eq:u1_tuning_gates}
\begin{aligned}
& u^{(1)}_n(\theta^{(v)}_n,\phi)=\\
    &\scalebox{0.7}{\begin{quantikz}
        & \gate[1]{P(\phi)}& & & &\gate[1]{P(\phi)} &\\
        & \ctrl{-1}  & \gate[1]{R_z(\pi/2)}& \gate[2]{XX+YY(\theta^{(v)}_{n})}&\gate[1]{R_z(-\pi/2)} & \ctrl{-1} &\\
        &  & & &  & &
    \end{quantikz}}
\end{aligned},
\end{equation}
where $P(\phi)=\text{diag}(1,e^{i\phi})$, and
\begin{equation} \label{eq:u3_tuning_gates}
\begin{aligned}
&u^{(3)}_n(\theta^{(v)}_n,\phi)=\\
    &\scalebox{0.7}{\begin{quantikz}
        &  & \gate[1]{R_z(\pi/2)}& \gate[2]{XX+YY(\theta^{(v)}_{n})}&\gate[1]{R_z(-\pi/2)} &  &\\
        & \ctrl{1} & & &  & \ctrl{1}& \\
        & \gate[1]{P(\phi)} & & &  & \gate[1]{P(\phi)}& \\
    \end{quantikz}}
\end{aligned},
\end{equation}
which define a tunable circuit family of unitaries $U_{\text{int}}(\theta^{(v)}_n,\phi)$ that depend on both $\theta^{(v)}_n$ and $\phi$.

This construction has two important limits. At $\phi=0$, the controlled-phase gates reduce to the identity and $U_{\text{int}}(0)$ generates non-interacting, free-fermion dynamics. Operator spreading may still occur in this limit, but the dynamics is not expected to exhibit the exponential growth associated with chaotic many-body scrambling. At $\phi=\pi$, the controlled-phase gates become controlled-$Z$ gates and the original strongly interacting chiral circuit $U_{\text{int}}$ is recovered. Intermediate values of $\phi$ continuously turn on the interaction-generating conditional phases, allowing us to track the emergence of scrambling through the OTOCs measured in Sec.~\ref{sec:scrambling}. Moreover, the total number of entangling controlled-$Z$ gates strongly depends on the chosen value of $\phi$ and the choice of boundary conditions. The actual circuit complexities for all the considered unitaries are presented in Appendix~\ref{Appendix: Circuit decompositions}.

\section{Hardware validation of semiclassical geometry}
\label{sec:Dispersion relation}

Having introduced the circuit realisations of the chiral model's semiclassical and interacting sectors, we now verify that the encoded dynamics reproduce the effective black hole geometry on superconducting quantum hardware. We focus on the semiclassical mean-field circuit and reconstruct its single-particle dispersion relation, which directly reveals the tilting of the effective light cone. By varying the relative homogeneous circuit couplings $\theta^{(v)}/\theta^{(u)}$, we resolve the transition from the exterior region to the horizon and into the over-tilted interior regime, thereby providing a hardware-level validation of the emergent spacetime geometry.

To reconstruct the single-particle dispersion of the mean-field circuit $U_{\rm MF}$, we utilise the spectroscopic protocol of Refs.~\cite{Roushan2017spectrum, Google2022boundstates}. A coherent superposition of the vacuum and a single excitation is prepared near the centre of the chain and evolved under $U_{\rm MF}$ with periodic boundary conditions. The amplitude and phase of the excitation relative to the vacuum are obtained by measuring the local coherence
\begin{equation} \label{eq:sigmaplus}
    \langle\sigma^{+}_{n}\rangle=\frac{1}{2}(\langle X_{n}\rangle + i\langle Y_{n}\rangle),
\end{equation}
on every qubit. A two-dimensional Fourier transform of the resulting spacetime signal $\langle\sigma^{+}_{n}(t)\rangle$ resolves the excitation's quasi-energy as a function of momentum, allowing the measured dispersion to be compared directly with Eq.~\eqref{eq:dispersion1}. Some of the values used for $\theta^{(v)}$ and $\theta^{(u)}$ in the hardware experiment are sufficiently large that the leading-order Trotter approximation is not guaranteed to remain accurate \textit{a priori}. However, direct numerical simulations of the corresponding Floquet circuits show that their single-particle spectra remain in good agreement with the Hamiltonian dispersion for the parameters considered here. This allows us to work at larger angles, thereby reducing the circuit depth required for the hardware measurement. Further details of the state preparation and spectroscopic reconstruction are given in Appendix~\ref{Appendix: Circuit decompositions}.

Figures~\ref{fig:overview}(c)-(e) show the measured dispersion for a fixed $\theta^{(u)}=\pi/4$ and varied $\theta^{(v)}=0$, $\pi/8$, and $\pi/4$, respectively. For $\theta^{(v)}=0$, the measured dispersion is untilted, corresponding to the exterior regime. At $\theta^{(v)}=\pi/8$, the chiral contribution tilts the dispersion and flattens one branch near the Fermi-point $k=-\pi/2$, identifying the horizon. At $\theta^{(v)}=\pi/4$, the dispersion becomes over-tilted, and additional Fermi-points appear, signalling the black hole interior. In all three regimes, the dominant hardware features agree with the Hamiltonian-limit prediction of Eq.~\eqref{eq:dispersion1}, providing a direct validation of the emergent light-cone structure on superconducting quantum hardware. These measurements were performed on \ibmboston{} for $N=24$ qubits and $M_{t}=20$ Trotter steps, using $10000$ shots per observable together with Pauli twirling, measurement readout twirling and zero-noise extrapolation. Further implementation and error-mitigation details are given in Appendix~\ref{Appendix:technical details}.

\section{Hawking temperature from coordinate-equivalent models}
\label{sec:Hawking temperature}

In this section, we probe Hawking thermality in three steps. We first show that the mean-field chiral and coordinate-equivalent XY Hamiltonians converge to the same Hawking temperature when their coupling profiles realise the same semiclassical geometry. We then use this equivalence to calibrate a local dynamical thermometer, relating the peak arrival time of an exterior population density pulse to the independently known Hawking temperature. Finally, we implement this protocol on superconducting quantum hardware, where the temperature can be estimated from a single-site population measurement without reconstructing the exterior density matrix.

\subsection{Coordinate-equivalent Hawking temperatures}

We first demonstrate that the mean-field chiral and XY Hamiltonians yield the same Hawking temperature when their coupling profiles are chosen to realise the same Schwarzschild spacetime geometry in the continuum. Both models belong to the common two-parameter family of Hamiltonians $H_{\rm MF}[u,v]$ defined in Eq.~\eqref{eq:MFHamiltonian}, where in general $u(x)$ and $v(x)$ can vary in space. The Schwarzschild geometry, expressed in Gullstrand-Painlevé coordinates, is realised in the chiral model's continuum limit by choosing $u_{\rm MF}=1$ and spatially varying $v_{\rm MF}(x)$, i.e.,
\begin{equation}
H_{\rm MF}[1,v_{\rm MF}].
\end{equation}
As shown in Sec.~\ref{sec:chiral}, the corresponding spacetime metric is that of Eq.~\eqref{eq:GPmetric}. The Schwarzschild geometry can also be expressed by the diagonal metric of Eq.~\eqref{eq:DiagonalMetricGeneral} and can be realised by the XY Hamiltonian, which is the special case
\begin{equation}
H_{\rm XY}[u_{\rm XY}] = H_{\rm MF}[u_{\rm XY},0],
\end{equation}
where the metric function $f(x)$ is related to the mean-field chiral Hamiltonian's coupling via
\begin{equation}
    f(x)=1-v_{\rm MF}(x)^2,
    \label{eq:metric_profile_matching}
\end{equation}
and the XY Hamiltonian's coupling via
\begin{equation}
    u_{\rm XY}(x)=f(x).
    \label{eq:XY_MF_matching}
\end{equation}
The two Hamiltonians are therefore microscopically distinct, but describe the same black hole geometry in their low-energy continuum limits \cite{sofos2025chiralentanglement}. In particular, they have the same surface gravity and Hawking temperature, which is determined by the gradient of the metric function at the horizon \cite{Volovik3};
\begin{equation}
    T_H=\frac{1}{4\pi}\at{\frac{df(x)}{dx}}{x=x_{h}}.
    \label{eq:hawking_temperature_profile}
\end{equation}
In the present paper, we choose the smooth metric profile
\begin{equation}
    f(x)=\frac{\pi}{2}\tanh\left[\alpha(x-x_{h})\right],
    \label{eqn:profile}
\end{equation}
for which the Hawking temperature is
\begin{equation}
    T_H=\frac{\alpha}{8}.
    \label{eq:hawking_temperature_alpha}
\end{equation}
For the comparison with the mean-field chiral model, only the near-horizon form of this profile is required. In this region $f(x)\ll1$, so the relation $f(x)=1-v_{\rm MF}(x)^2$ admits a real $v_{\rm MF}(x)$ and the two descriptions can be matched locally. Since the Hawking temperature depends only on the near-horizon gradient $f'(x_{h})$, no global matching of the full profile is required.
\begin{figure*}
    \centering
\includegraphics[width=0.99\linewidth]{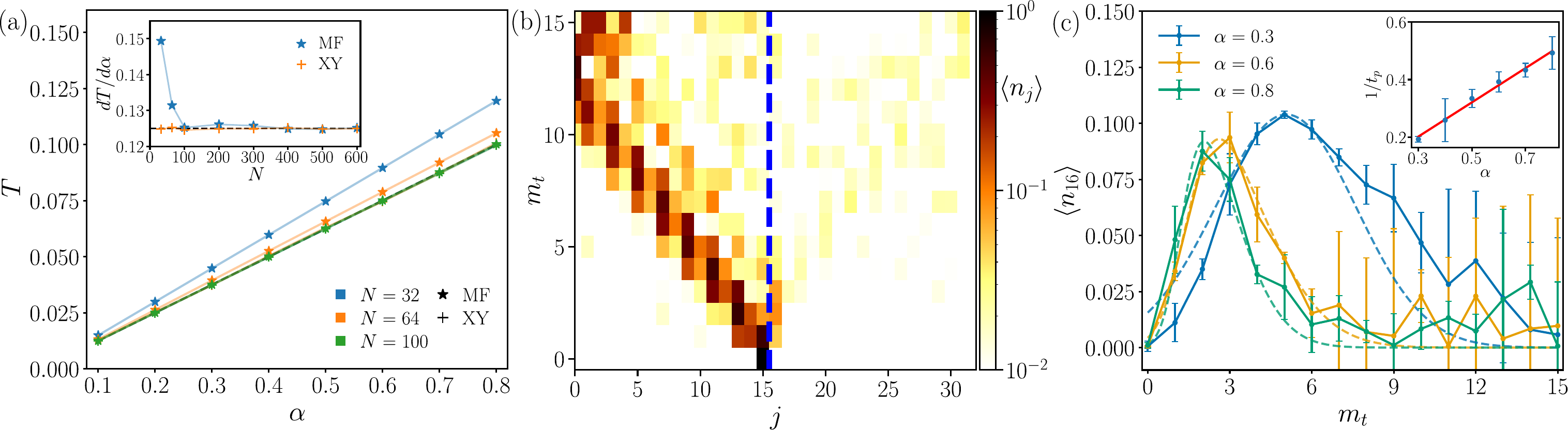}
\caption{Extraction of the Hawking temperature in the semiclassical black hole simulator.
(a) Temperature obtained by fitting the exterior mode occupations of the XY and mean-field chiral Hamiltonians to the thermal distribution in Eq.~\eqref{eq:Hawking probabilities}, shown as a function of the profile parameter $\alpha$ for different system sizes $N$. The black dashed line denotes the continuum prediction for the Hawking temperature given by Eq.~\eqref{eq:hawking_temperature_alpha}. Inset: finite-size scaling of the derivative $dT/d\alpha$ for the two models, demonstrating convergence towards the theoretical value.
(b) Site-resolved occupation density during the evolution of a wave packet under $U_{\rm XY}$ on \ibmboston{}, for $N=32$, $\alpha=0.8$, and the inhomogeneous profile in Eq.~\eqref{eqn:profile}. The excitation is initially localised at site $j=15$, and the dashed vertical line marks the horizon at $n_{h}=15.5$. (c) Mitigated hardware-measured occupation of the exterior site $j=16$ for different values of $\alpha$. The dashed curves are skew-Gaussian fits used to determine the peak time $t_p$. Inset: extracted inverse peak time $t_p^{-1}$ plotted against $\alpha$. The red line shows the expected relation with the ansatz $t_p^{-1}=\alpha/(8\gamma)$, with $\gamma\approx0.197\pm0.008$. Error bars on the density measurements represent the uncertainty of the mitigated hardware data, while those in the inset are propagated from the fitted peak-time uncertainties according to $\sigma_{1/t_p}=\sigma_{t_p}/t_p^2$.}
\label{fig:XY_hawking}
\end{figure*}

Hawking emission from spatially inhomogeneous lattice and Floquet models has likewise been analysed directly at the level of the outgoing spectrum~\cite{Maertens_2024}. Operationally, we first extract the Hawking temperature for the two Hamiltonians numerically using an energy-resolved quench protocol. Following an initial quench, the state $|\psi(0)\rangle$ is time-evolved under either $H_{\rm XY}$ or $H_{\rm MF}$ to obtain $|\psi(t)\rangle$, allowing it to partially tunnel outside the black hole~\cite{benhemou2025opticalblackhole, HornerHallamPachos2023,yang2020simulating}. For an observer restricted to the exterior, the time-evolved state is resolved in the eigenbasis of the exterior Hamiltonian $H_{\rm out}$. Thus, the probability $P_m$ of the state occupying an eigenstate $|E_m\rangle$ of $H_{\rm out}$ with energy $E_m$ is $P_m = \langle E_m|\rho_{\rm out}|E_m\rangle$ \cite{benhemou2025opticalblackhole}, where $\rho_{\rm out}=\rm{Tr}_{\rm in}\rho(t)$ is the density matrix of the time-evolved state $\rho(t)=|\psi(t)\rangle\langle\psi(t)|$ with the degrees of freedom inside the horizon traced out, and $H_{\rm out}$ is taken to be either $H_{\rm XY}$ or $H_{\rm MF}$ defined over only the exterior subsystem. In the Hamiltonian's low-energy limit, we find that this occupation probability follows a thermal Boltzmann distribution,
\begin{equation}
P_m \propto e^{-E_m/T_H},
\label{eq:Hawking probabilities}
\end{equation}
allowing $T_H$ to be extracted from the gradient of $\log (P_m)$ as a function of $E_m$~\cite{gibbon_qgravity}.

Figure~\ref{fig:XY_hawking}(a) shows the temperatures extracted from this distribution for the XY and mean-field chiral models as the system size is increased. For both Hamiltonians, we take a near-horizon approximation of $f(x)$ such that the XY coupling is $u_{\rm XY}(x)= \frac{\pi\alpha}{2}(x-x_{h}) + \mathcal{O}(x^3)$ with $x_h=a(N-1)/2$ and the mean-field coupling is $v_{\rm MF}(x)=1-\frac{\pi\alpha}{4}(x-x_{h})+\mathcal{O}(x^2)$, where $x_{h}=aN/2$. In the former case, the horizon is placed midway between two lattice sites to avoid the tunnelling amplitude vanishing at a certain position~\cite{benhemou2025opticalblackhole, yang2020simulating}. The initial state $|\psi(0)\rangle$ is taken to be a single-particle excitation localised behind the horizon at site $j=0.475N$, rounded to the nearest integer, and is evolved until the time $t_p$ at which the local population $n_j(t) = \langle c_j^\dagger c_j\rangle$ of the exterior site adjacent to the horizon peaks. Despite their different microscopic couplings, we see that both models converge to the same continuum value $T_H=\alpha/8$, confirming their equivalence at the level of semiclassical horizon thermality. The XY Hamiltonian exhibits smaller finite-size corrections and converges more rapidly to the asymptotic result. Together with its substantially lower circuit depth, this makes the XY realisation the natural choice for the hardware implementation below.

\subsection{Numerical calibration of a local dynamical thermometer}

Although the energy-resolved protocol provides a direct numerical test of Hawking thermality, implementing it on current quantum hardware would require reconstructing the exterior's reduced density matrix and resolving it in the eigenbasis of $H_{\rm out}$. We therefore introduce a hardware-compatible dynamical thermometer that relies solely on the time-dependent population of a single site near the horizon. 

Following Ref.~\cite{benhemou2025opticalblackhole}, we prepare $|\psi(0)\rangle$ to be a single-particle excitation localised just behind the horizon $x_{h}=a(N-1)/2$ and evolve it under the inhomogeneous XY circuit with the coupling $u_{\rm XY}(x)=f(x)$. As the excitation propagates through the near-horizon regime, we observe a pulse in the local population $n_j$ of the exterior site adjacent to the horizon. For the family of near-horizon profiles considered here, the characteristic peak time $t_p$ is predicted to scale inversely with the surface gravity and, hence, with the Hawking temperature~\cite{benhemou2025opticalblackhole}. We therefore define the calibrated estimator
\begin{equation}
    T_H^{\rm(est)}=\frac{\gamma}{t_p},
    \label{eq:dynamical_thermometer}
\end{equation}
where $\gamma$ is a protocol-dependent calibration constant. For a Trotterised implementation, $\gamma$ depends on the discrete-time protocol, including the system size, initial and monitored sites, and finite time step, and approaches the corresponding Hamiltonian value as $\delta t\rightarrow 0$.

We determine $\gamma$ by applying the protocol to reference profiles whose temperatures are known theoretically from Eq.~\eqref{eq:hawking_temperature_alpha}. A fit of $t_p^{-1}$ against $T_H$ then calibrates the local population measurement as a dynamical thermometer. Once calibrated, the same single-site measurement can be used to estimate the Hawking temperature of another profile within the same calibrated family and experimental configuration. Details of this calibration for the present circuit geometry are provided in Appendix~\ref{sec:appendix_ED}, where fitting the reference data gives $\gamma\approx0.186$.

\subsection{Hardware implementation of local dynamical thermometer} 

We implement the dynamical thermometer on a chain of $N=32$ qubits, placing the horizon at $n_{h}=(N-1)/2$ and preparing the initial excitation at $j=N/2-1$. In the Floquet implementation, the corresponding XY rotation angles are related to the Hamiltonian couplings by $\theta_n^{(v)}=0$ and $\theta_n^{(u)} = -u_{\rm XY}(x_n)\delta t$, giving inhomogeneous rotations
\begin{equation}
    \theta_n^{(u)}=-f(n),
\end{equation}
where we have fixed $a=1$ and set $\delta t=1$ for this Hawking temperature experiment. The site-resolved population is obtained directly from the Pauli-$Z$ expectation value,
\begin{equation}
    \langle n_j(t)\rangle =\frac{1-\langle Z_j(t)\rangle}{2}.
\end{equation}

Figure~\ref{fig:XY_hawking}(b) shows the hardware-measured propagation for $N=32$ and $\alpha=0.8$ after we have performed the error mitigation strategy defined in Appendix~\ref{app:TH_decoherence}. The dashed line marks the horizon located at $n_h=15.5$. The dominant component of the wave packet remains on the interior side, while a non-zero density response develops on the exterior side and is detected by monitoring the sites adjacent to the horizon. Figure~\ref{fig:XY_hawking}(c) shows the population measured at site $j=16$ for several values of $\alpha$. Increasing $\alpha$ steepens the near-horizon metric function, increasing the Hawking temperature and causing the exterior population peak to occur at earlier times. 

For each value of $\alpha$, we fit the exterior population pulse with a skew-Gaussian and extract its peak time $t_p$. The inset of Fig.~\ref{fig:XY_hawking}(c) shows $t_p^{-1}$ against $\alpha$, which is related independently to the Hawking temperature through $T_H=\alpha/8$. The approximately linear dependence verifies the predicted scaling $t_p\propto T_H^{-1}$ and calibrates Eq.~\eqref{eq:dynamical_thermometer}, giving $\gamma\simeq0.197\pm0.008$ for the present experimental configuration. The experimentally calibrated value is consistent with the exact-diagonalisation result $\gamma\simeq0.186$ obtained in Appendix~\ref{sec:appendix_ED}, providing a non-trivial validation of the dynamical-thermometer protocol on quantum hardware.

The hardware measurement should therefore be viewed as a calibrated dynamical thermometer: once $\gamma$ is fixed from reference profiles, the Hawking temperature of another profile within the same calibrated family and experimental configuration can be estimated from the peak time of a single exterior-site population. Moreover, the agreement between the experimental and exact-diagonalisation values of $\gamma$ shows that the hardware faithfully reproduces the near-horizon dynamics. Together with the numerical demonstration in Fig.~\ref{fig:XY_hawking}(a) that the corresponding semiclassical evolution produces thermal exterior occupations, this provides evidence that the hardware-observed exterior response is consistent with the expected Hawking-thermal dynamics. Establishing its thermal character directly on hardware would, however, require resolving the exterior occupation probabilities in energy, which is considerably more experimentally demanding.

All hardware measurements were performed on \ibmboston{} with $10000$ shots per observable and using the error-mitigation procedures detailed in Appendix~\ref{Appendix:technical details}. We simulated $N=32$ qubits for up to $M_{t}=15$ Trotter steps, measuring the site populations after each step. For more information on the technical details, error mitigation methods and $2$-qubit gate depths, see Appendix~\ref{Appendix:technical details}.

\section{Tunable scrambling and Lyapunov growth}
\label{sec:scrambling}

Having verified the semiclassical horizon dynamics, we now turn to the interacting sector of the chiral model responsible for many-body scrambling. Building on Refs.~\cite{Daniel2025chiralscrambling, daniel2025blackholeteleport}, we use the tunable gate decompositions in Eqs.~\eqref{eq:u1_tuning_gates} and~\eqref{eq:u3_tuning_gates} to interpolate continuously between the free-fermion limit at $\phi=0$ and the strongly interacting chiral regime at $\phi=\pi$. We measure out-of-time-ordered correlators on superconducting quantum hardware and determine how Lyapunov growth emerges as the interactions are switched on.

\subsection{OTOCs and Lyapunov growth}

A standard diagnostic of operator spreading and quantum chaos is the out-of-time-ordered correlator (OTOC)~\cite{scramblingandlyapunov, Kukuljan_2017,hoshur_chaos, Swingle2016OTOC}. At infinite temperature, we consider
\begin{equation}
C(t) = \frac{1}{2^N} \mathrm{Tr}\left[ B(t) M B(t) M \right],
\label{eq:OTOC}
\end{equation}
where $B(t)=U^\dagger(t)BU(t)$ is a local butterfly operator evolved in the Heisenberg picture and $M$ is a local measurement operator~\cite{Google2024OTOC,Google2025OTOC}. When $B$ and $M$ initially commute, the decay of $C(t)$ measures the extent to which the time-evolved operator $B(t)$ develops support at the position of $M$, and therefore probes the spreading of quantum information through the system. This quantity is the conventional OTOC widely used to diagnose operator growth and information scrambling in many-body systems~\cite{Google2024OTOC,Braumuller2022OTOCexperiment, Nahum2018OTOCcircuits, Jensen_2016,Maldacena_2016, Hashimoto_2017}.

In chaotic interacting systems, the early-time OTOC decay can exhibit Lyapunov growth. We define
\begin{equation}
F_N(t)
=
U\!\left(
\frac{1}{2},1,Ne^{-\lambda_L t}
\right)
\sqrt{N}\,e^{-\lambda_L t/2},
\end{equation}
where $U$ is Kummer's confluent hypergeometric function. As the measured correlator is normalised so that $C(0)=1$, we use the finite-size fitting function
\begin{equation}
C_{\rm fit}(t)
=
\frac{F_N(t)}{F_N(0)}.
\label{eq:kummer}
\end{equation}
This form captures finite-size corrections to the early-time scrambling regime and has previously been applied to strongly chaotic models such as the Sachdev-Ye-Kitaev model~\cite{SYK_chaos}.

Although the universal chaos bound is conventionally formulated at finite temperature~\cite{MaldacenaShenkerStanford2016}, here we work at infinite temperature, which is directly accessible through the random-state echo protocol implemented below. We compare the hardware extracted decay rate with the infinite-temperature Lyapunov exponent of $\lambda_L/v \simeq0.78$ previously obtained for the $N=12$ interacting chiral spin-chain~\cite{Daniel2025chiralscrambling, daniel2025blackholeteleport}.

The interpretation of the fit depends crucially on the interaction strength. In the strongly interacting chiral regime, $\phi=\pi$, the OTOC is expected to display an early-time exponential decay, so that Eq.~\eqref{eq:kummer} yields a meaningful chaotic Lyapunov exponent. By contrast, at the free-fermion point $\phi=0$, the OTOC may decay through ballistic operator spreading, dephasing and finite-size interference, but this decay is not expected to be genuinely exponential. A fit to Eq.~\eqref{eq:kummer} in this regime therefore defines only an effective decay scale rather than a true Lyapunov exponent. The tunable parameter $\phi$ consequently allows us to probe the emergence of Lyapunov growth as interactions are introduced.

\begin{figure*}
    \centering
    \includegraphics[width=0.99\linewidth]{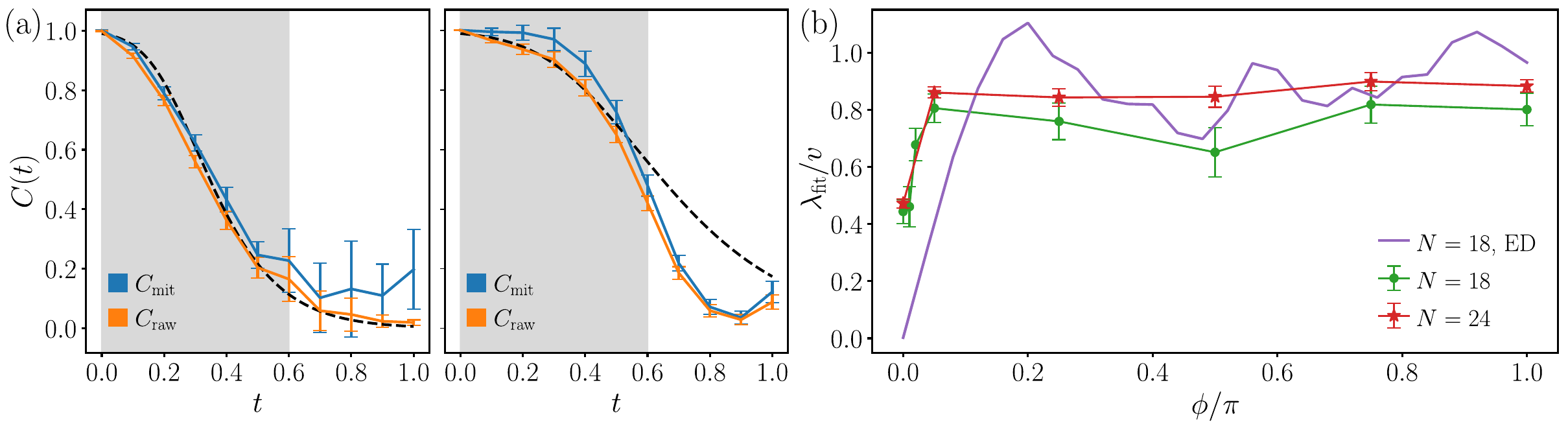}
\caption{Tunable scrambling measured on superconducting quantum hardware. 
(a) Left: OTOC $C(t)$ in the strongly interacting chiral regime, $\phi=\pi$, for $N=24$ spins. Blue and orange points show the error-mitigated and raw hardware data, respectively. The black dashed curve is a fit to Eq.~\eqref{eq:kummer}, and the shaded region indicates the fitting window [$\chi^2_\nu\approx1.26$].  
(a) Right: Corresponding OTOC dynamics in the free-fermion limit, $\phi=0$. Here, the Lyapunov form gives a poor description of the decay, and the fitted value should be interpreted only as an effective decay scale [$\chi^2_\nu\approx2.50$]. 
(b): Fitted decay scale as a function of $\phi$ for $N=18$ and $N=24$ spins. For sufficiently strong interactions, the extracted values are close to the theoretical Lyapunov exponent $\lambda_L/v\simeq0.78$ of the $N=12$ chiral spin model at infinite temperature found using ED~\cite{Daniel2025chiralscrambling, daniel2025blackholeteleport}. All hardware simulations were performed on \ibmboston{} using $10000$ shots per circuit and averaged over 20 random initial bitstrings. As a baseline, we also include state-vector simulations for $N=18$ spins, averaged over 500 initial bitstrings to approximate the infinite-temperature trace.}
\label{fig:hardware scrambling}
\end{figure*}

For the measurements below, we choose $M=X_i$ and $B=X_j$, with $i=N/2$ and $j=N/2-2$ on an open chain of $N=18$ and $N=24$ spins. Placing both operators near the centre suppresses boundary effects, while their short separation allows the onset of operator spreading to be resolved within the coherence time available on current hardware.
\subsection{Tunable echo protocol and error mitigation}

To measure the OTOC defined in Eq.~\eqref{eq:OTOC}, we implement a hardware-compatible echo protocol based on Refs.~\cite{Google2024OTOC, Google2025OTOC}. For a fixed value of $\phi$, we denote the evolution over $m_{t}$ Trotter steps as:

\begin{equation}
    U_{\phi}(t)
    =
    \left[U_{\rm int}(\theta,\phi)\right]^{m_t},
    \qquad
    t=m_t\delta t,
\end{equation}
where $U_{\phi}(t)$ is decomposed according to Eq.~\eqref{eq:int_gate_structure} and the local three-qubit unitaries given by Eqs.~\eqref{eq:u1_tuning_gates} and~\eqref{eq:u3_tuning_gates}.
At $\phi=0$, the controlled-phase gates reduce to the identity and the circuit generates free-fermion dynamics, whereas at $\phi=\pi$ they become controlled-$Z$ gates and recover the strongly interacting chiral evolution.

For each measurement, the qubit at the position of the probe operator $M$ is prepared in its $+1$ eigenstate, while all remaining qubits are initialised in a randomly sampled computational-basis state. The circuit applies the forward evolution $U_{\phi}(t)$, the local butterfly operator $B$, and the reversed evolution $U_{\phi}^{\dagger}(t)$, before measuring $M$. Averaging the resulting echo expectation value over random initial bitstrings provides a stochastic estimate of the infinite-temperature trace in Eq.~\eqref{eq:OTOC}. In the experiments below, both $M$ and $B$ are Pauli $X$ operators. The complete circuit and the relation between the random-state average and the infinite-temperature OTOC are given in Appendix~\ref{Appendix: Circuit decompositions}.

In addition to Pauli twirling, measurement-readout twirling, and zero-noise extrapolation, we employ an echo-specific normalisation following Ref.~\cite{Google2024OTOC}. We execute a reference circuit that is identical to the OTOC circuit except that the butterfly operator $B$ is omitted. Ideally, the forward and backward evolutions then cancel exactly, and the measured probe expectation value remains equal to unity. Any decay of this reference signal, therefore, quantifies the attenuation caused by decoherence, gate errors and imperfect reversal.

Since $B$ is a single-qubit Pauli gate, its contribution to the total circuit error is small compared with that of the forward and backward many-body evolutions. Consequently, the OTOC and reference circuits experience approximately the same background attenuation. We correct for this common-mode decay by dividing the measured OTOC by the corresponding reference signal and normalising the result so that $C(0)=1$. This procedure suppresses hardware-induced decay while retaining the additional loss of the echo generated by operator spreading. The precise normalisation and reference circuits are described in Appendix~\ref{Appendix: Circuit decompositions}.

\subsection{Hardware observation of tunable scrambling}

We implement the OTOC protocol on \ibmboston{} for open chains of $N=18$ and $24$ qubits and evolution times of up to $M_t=10$ Trotter steps. Each observable is estimated using $10000$ shots and averaged over 20 randomly sampled initial bitstrings. The raw hardware data is processed using the error-mitigation procedures described in the preceding subsection.



The left panel of Fig.~\ref{fig:hardware scrambling}(a) shows the measured OTOC in the fully interacting chiral regime, $\phi=\pi$. We use spatially uniform circuit parameters with
\begin{equation}
    \theta^{(v)}
    =
    \frac{2\delta t\,v}{16},
    \qquad
    \delta t=0.1,
    \qquad
    v=20.
\label{eq:theta}
\end{equation}
We note that the $1/16$ scaling in $\theta^{(v)}$ is chosen to agree with the Hamiltonian dynamics performed in Ref.~\cite{Daniel2025chiralscrambling, daniel2025blackholeteleport}. Both the raw and error-mitigated data exhibit a rapid initial decay. Fitting the mitigated correlator over the interval $0\leq t\leq0.6$ using Eq.~\eqref{eq:kummer} gives
\begin{equation}
    \lambda_{\rm fit}/v=0.88\pm0.02.
\end{equation}
Since the OTOC protocol evaluates the infinite-temperature trace, this may be compared with the corresponding infinite-temperature value $\lambda_L/v\simeq0.78$ previously obtained numerically for the $N=12$ interacting chiral spin-chain Hamiltonian~\cite{Daniel2025chiralscrambling, daniel2025blackholeteleport}. Because Eq.~\eqref{eq:kummer} explicitly incorporates finite-size corrections and the available Hamiltonian benchmark $\lambda_L/v\simeq0.78$ was obtained for $N=12$, this value should be regarded as a reference scrambling scale rather than an exact prediction for the $N=24$ experiment. Although the experimental value lies above the numerical result, the two reproduce the same characteristic scrambling scale. The agreement shows that the reduced Trotterised circuit retains the characteristic Lyapunov growth of the strongly scrambling chiral model.

The right panel of Fig.~\ref{fig:hardware scrambling}(a) shows the corresponding result at $\phi=0$. At this point, the controlled-phase gates reduce to the identity, and the circuit becomes a free-fermion evolution. The OTOC still decays because the initially local operator spreads ballistically through the chain and undergoes finite-size dephasing and interference. Its time dependence, however, is not well described by the Lyapunov form in Eq.~\eqref{eq:kummer}. A formal fit gives
\begin{equation}
    \lambda_{\rm fit}/v=0.47\pm0.02,
\end{equation}
but this quantity should be interpreted only as an effective decay scale and not as a chaotic Lyapunov exponent, as demonstrated by the relatively poor quality of the fit.

This non-exponential behaviour is not caused by interaction-generating Trotter corrections. At $\phi=0$, the circuit is generated entirely by operators quadratic in the Jordan-Wigner fermions, and such operators remain quadratic under commutation. Trotter corrections can therefore modify the effective free-fermion dynamics, for example, by generating longer-range hopping, but cannot induce genuine many-body interactions. The remaining deviations are instead attributable to finite-size effects, finite sampling and hardware noise.

Finally, Fig.~\ref{fig:hardware scrambling}(b) shows the fitted decay scale as the controlled-phase angle is varied between the free and interacting limits for $N=18$ and $N=24$. As a comparison, we also include state-vector simulations for $N=18$ spins, averaged over 500 initial bitstrings to approximate the infinite-temperature trace. At very small phases, such as $\phi=0.01\pi$, the circuit remains perturbatively close to the free-fermion point and the fitted value remains close to the corresponding effective decay scale. As $\phi$ is increased, the OTOC becomes increasingly well described by the Lyapunov form, and the extracted values for sufficiently strong interactions are close to the theoretical chiral spin-chain model prediction.

As a benchmark, exact diagonalisation of the chiral spin-chain Hamiltonian at infinite temperature gives $\lambda_L/v\simeq0.78$ for $N=12$~\cite{Daniel2025chiralscrambling, daniel2025blackholeteleport}. For the larger systems considered here, state-vector simulations of the Trotterised circuit give $\lambda_L/v\simeq0.90$ for $N=18$, while the corresponding hardware measurements yield $\lambda_{\rm fit}/v\simeq0.80$ for $N=18$ and $\lambda_{\rm fit}/v\simeq0.88$ for $N=24$. These values remain close to the original chiral-model result and to one another, extending the numerical benchmark to larger system sizes and demonstrating that the Trotterised hardware implementation retains the characteristic scrambling scale of the interacting chiral model.

Increasing the number of random initial bitstrings would reduce the statistical uncertainty associated with approximating the infinite-temperature trace. Improvements in circuit depth, coherence and error mitigation would additionally extend the accessible fitting window and sharpen the distinction between the non-exponential free regime and the interacting Lyapunov regime.

\section{Conclusions}

We have demonstrated a superconducting-qubit simulation of two central dynamical aspects of black hole physics: Hawking thermality and interior scrambling. Starting from a chiral spin-chain model with an emergent horizon, we constructed hardware-compatible Floquet circuits that separate the semiclassical free-fermion sector from the interacting chiral sector. This allowed us to use lower-depth circuits to probe the effective spacetime geometry and Hawking temperature, while retaining an interacting circuit capable of generating many-body scrambling.

In the semiclassical sector, we measured the single-particle dispersion relation of the mean-field circuit on IBM hardware. The measured spectra reproduce the progression from an untilted exterior dispersion, through the horizon condition with a flattened branch, to an over-tilted interior dispersion, directly resolving the effective light-cone structure. We also showed numerically that the XY and mean-field chiral models converge to the same Hawking temperature when their coupling profiles describe the same effective metric. The lower-depth XY circuit was then used to monitor the propagation of a localised excitation near the horizon, where the inverse relation between the exterior pulse-arrival time and the gradient of the coupling profile provides a local dynamical estimator of the Hawking temperature.

Beyond the semiclassical regime, we introduced a tunable interacting circuit that continuously interpolates between free-fermion and strongly interacting chiral dynamics via the controlled-phase angle $\phi$. Using an infinite-temperature OTOC echo protocol, we distinguish non-exponential operator spreading at $\phi=0$ from Lyapunov-like scrambling in the interacting regime. At the free-fermion point, the fitted value $\lambda_{\rm fit}/v\simeq0.47$ represents only an effective decay scale rather than a genuine chaotic Lyapunov exponent, whereas a well-defined Lyapunov-like decay emerges as the interaction-generating phase $\phi$ is increased.

Across these complementary probes, the hardware results remain quantitatively close to the corresponding analytical and numerical benchmarks. The measured dispersion relations closely reproduce the analytical band structure across the exterior, horizon and interior regimes. For Hawking thermality, the experimentally calibrated dynamical thermometer yields a value of $\gamma$ consistent with the exact-diagonalisation result $\gamma=0.186$. In the interacting regime, the measured scrambling rates $\lambda_{\rm fit}/v\simeq0.80$ for $N=18$ and $\lambda_{\rm fit}/v\simeq0.88$ for $N=24$ remain close to the infinite-temperature chiral-model result $\lambda_L/v\simeq0.78$ obtained for $N=12$, as well as to the larger-system Trotterised value $\lambda_L/v\simeq0.90$ for $N=18$. These measurements, therefore, extend the scrambling benchmark to larger system sizes while retaining the characteristic Lyapunov scale of the chiral model.

Taken together, these results establish a programmable quantum-simulation framework in which semiclassical horizon thermality, coordinate-equivalent black hole geometries, and interacting interior scrambling can be studied on the same hardware platform. The observable-specific circuit reductions allow these distinct regimes to be accessed within a common microscopic framework while remaining compatible with the circuit depths available on present quantum processors. This provides a route towards studies of information-recovery protocols, Page-curve diagnostics, Hayden-Preskill decoding, and interaction-induced corrections to semiclassical Hawking radiation. Extending the chiral model and its Floquet circuit beyond one dimension would further enable the investigation of higher-dimensional horizon geometries, spatially resolved scrambling, and, with suitable extensions, rotating black hole analogues.

Future improvements in coherence, connectivity, circuit compilation, and error mitigation should enable progressively more complete implementations of the parent chiral evolution. This would allow horizon thermality and interacting scrambling to be studied simultaneously within a single spatially inhomogeneous circuit and at system sizes for which generic interacting real-time dynamics becomes increasingly demanding to simulate classically.
\begin{acknowledgments}
We thank Matthew Yusuf, Aiden Daniel, and Tanmay Bhore for their valuable discussions. All circuit simulations were performed using the Qiskit SDK \cite{Qiskit2024}. Computational portions of this research were carried out on ARC4 and AIRE, part of the High-Performance Computing facilities at the University of Leeds. This project was funded and supported by the UK National Quantum Computer Centre [NQCC200921], which is a UKRI Centre and part of the UK National Quantum Technologies Programme (NQTP). R.S. acknowledges support by the Leverhulme Trust Research Leadership Award RL-2019-015 and EPSRC Grants No. EP/Z533634/1 and No. UKRI1337. I.A.S. acknowledges support from EPSRC with Grant No. EP/W524372/1. 

\end{acknowledgments}

{\bf Data Availability:}
Data will become available upon reasonable request.

\newpage
\appendix

\section{Derivation of the chiral mean-field Floquet model}\label{Appendix:JW mean field}

Sec.~\ref {subsec:floquet_chiral} introduces a Floquet version of the chiral black hole simulator that can be run on quantum hardware. Depending on the observables measured, two simplifications can be implemented to reduce the complexity of this Floquet model. First is an approximation we use when simulating the scrambling dynamics of black holes at reduced circuit depth. Second is a Floquet version of a mean-field Hamiltonian that describes the spacetime curvature in the semi-classical limit. This mean-field Floquet operator contains the information of the tilting dispersion relation for homogeneous couplings, and the thermalisation of wave packets to the Hawking temperature for spatially varying couplings (see Sec.\ref{sec:Dispersion relation} and \ref{sec:Hawking temperature} respectively). 
In this appendix, we explicitly derive the Floquet operator of the chiral model's mean-field limit.

We begin with the definition of the chiral Hamiltonian of Eq.\eqref{eq:chiralHamiltonian} that, using the spin raising and lowering operators $\sigma^\pm_n=(X_n\pm iY_n)/2$, can be expressed as \cite{HornerHallamPachos2023, Forbes2023interactingchiralmodel}
\begin{equation}
\begin{split}
    H=-\frac{1}{a}&\sum_{n=1}^N\Bigl[ u_n\sigma^+_n\sigma_{n+1}^- -\frac{iv_n}{2}\bigl(\sigma^+_n\sigma^-_{n+1}Z_{n+2} \\ & +\sigma^+_{n+1}\sigma^-_{n+2}Z_{n}+\sigma^+_{n+2}\sigma^-_nZ_{n+1}\bigr)\Bigr] +\text{H.c.}
    \label{Equation: Chiral Hamiltonian (Spin Ladder Operators)}
\end{split}
\end{equation}
The spin Hamiltonian of Eq.~\eqref{Equation: Chiral Hamiltonian (Spin Ladder Operators)} can be mapped to a fermionic Hamiltonian via the Jordan-Wigner transformation:
\begin{equation}
    \sigma^+_n=\exp(-i\pi\sum_{i<n}c^\dagger_ic_i)c^\dagger_n, \quad \sigma^-_n=\exp(i\pi\sum_{i<n}c^\dagger_ic_i)c_n,
\end{equation}
where $Z_n=(1-2c^\dagger_nc_n)=\exp(\pm i\pi c^\dagger_n c_n)$, such that $\sigma_n^+\sigma^-_{n+1}=c^\dagger_nc_{n+1}$ and $\sigma_{n+2}^+\sigma^-_n=c^\dagger_{n+2}\sigma^z_{n+1}c_n$. Doing so gives
\begin{equation}
\begin{split}
    H=-\frac{1}{a}\sum_{n=1}^N\Bigl[& u_nc^\dagger_nc_{n+1} -\frac{iv_n}{2}\bigl(c^\dagger_nc_{n+1}Z_{n+2} \\ & \ \ +c^\dagger_{n+1}c_{n+2}Z_{n}-c^\dagger_{n}c_{n+2}\bigr)\Bigr] +\text{H.c.},
    \label{Equation: Chiral Hamiltonian (Fermionic Operators)}
\end{split}
\end{equation}
where $c_n^\dagger$ and $c_n$ are fermionic creation and annihilation operators that satisfy $\{c^\dagger_n,c_m\}=\delta_{nm}$ with all other anticommutation relations vanishing.

Expressing the chiral Hamiltonian in terms of fermionic operators makes it straightforward to apply the mean-field approximation. To do so, we replace the operators $Z_n$ with their expectation value with respect to the mean-field Hamiltonian's ground state; that is, $Z_n\mapsto\langle Z \rangle$, where we have assumed translational invariance and dropped the site index $n$ \cite{HornerHallamPachos2023, Forbes2023interactingchiralmodel}. As the chiral Hamiltonian is particle-hole symmetric, it can be inferred that its ground state is half-filled and, therefore, $\langle Z\rangle=0$ \cite{HornerHallamPachos2023, Forbes2023interactingchiralmodel}. It follows that the mean-field limit of the chiral Hamiltonian is
\begin{equation}
    H_{\rm MF}=-\frac{1}{a}\sum_{n=1}^N\left( u_nc^\dagger_nc_{n+1} +\frac{iv_n}{2}c^\dagger_nc_{n+2}\right)+\text{H.c.},
\end{equation}
which, after transforming back to spin operators, is equivalent to
\begin{align}
    H_{\rm MF}&=-\frac{1}{a}\sum_{n=1}^N\left( u_n\sigma^+_n\sigma^-_{n+1}+\frac{iv_n}{2}\sigma^+_nZ_{n+1}\sigma^-_{n+2}\right)+\text{H.c.}, \\
    \begin{split}
    &=\frac{1}{a}\sum_{n=1}^N\biggl[ -\frac{u_n}{2}(X_nX_{n+1}+Y_nY_{n+1}) \\ 
                & \quad \quad \quad \quad + \frac{v_n}{4}(Y_nZ_{n+1}X_{n+2}-X_nZ_{n+1}Y_{n+2})\biggr].
    \end{split}
    \label{eq:Mean field JW}
\end{align}

Finally, as performed in Sec.~\ref{subsec:floquet_chiral}, we consider a first-order Trotterisation of the time evolution operator governed by Eq.\eqref{eq:Mean field JW} to obtain the mean-field Floquet operator
\begin{equation}
\begin{aligned}
U_{\mathrm{MF}}&=\exp\Bigg[-i\sum_{n=1}^{N}\frac{\theta^{(v)}_{n}}{2}    \big(Y_{n}Z_{n+1}X_{n+2}-X_{n}Z_{n+1}Y_{n+2}\big)\Bigg]\\
                &\quad \times\exp\Bigg[-i\sum_{n=1}^{N}\frac{\theta^{(u)}_{n}}{2} \big(X_{n}X_{n+1}+Y_{n}Y_{n+1}\big)\Bigg]
\end{aligned}
\end{equation}
where $\theta^{(u)}_n=-u_n\delta t$ and $\theta^{(v)}_n=v_n\delta t/2$ encode the $u_n$ and $v_n$ couplings, respectively. We observe that this is the same model as given in Eq.~\eqref{eq:MFfloquetcircuit}, where we simply neglect the interacting terms from the full chiral model. As shown in Sec.~\ref{subsec: chiral circuit encoding}, this Floquet operator can be further decomposed for execution on quantum hardware.

\section{Circuit decompositions}\label{Appendix: Circuit decompositions}

\subsection{Decomposition of the chiral Floquet unitary}
\label{Appendix:chiral Floquet decomposition}

A direct hardware-efficient implementation of the full three-spin evolution $\exp(-i\theta\chi_n)$ is not known for an arbitrary rotation angle $\theta$. A special case has been identified in Ref.~\cite{Reascos2023chiralitycircuits}: for the rescaled chirality operator $4\chi_n/\sqrt{3}$ and the fixed angle $\theta=2\pi/3$, the resulting unitary can be implemented using two SWAP gates. This special point is nevertheless free fermionic and does not generate the general interacting chiral dynamics required for the scrambling protocol considered here.

Using the decomposition of the chiral spin-chain Hamiltonian into the local Hamiltonians of Eq.~\eqref{Equation: Local Hamiltonians}, and the definition of the local unitaries $u_n^{(i)}(\theta)$ given in Eq.~\eqref{Equation: Local Trotterisation Unitaries}, we apply a first-order Trotter decomposition to the three components of the chirality operator. This gives the parent chiral Floquet unitary
\begin{equation}\label{eq:fullunitarybreakdown}
\begin{aligned}
U_{\rm C}
={}&
\exp\Big(-i\sum_{n}^{N}\frac{\theta_n^{(v)}}{2}h^{(3)}_{n}\Big)
\exp\Big(-i\sum_{n}^{N}\frac{\theta_n^{(v)}}{2}h^{(2)}_{n}\Big)
\\
&\times
\exp\Big(-i\sum_{n}^{N}\frac{\theta_n^{(v)}}{2}h^{(1)}_{n}\Big)
\exp\Big(-i\sum_{n}^{N}\frac{\theta_n^{(u)}}{2}h^{(0)}_{n}\Big),
\end{aligned}
\end{equation}
This unitary serves as the parent Floquet model from which the mean-field, XY, and interacting circuit families used in the main text are obtained by retaining the components relevant to each physical protocol. The additional Trotterisation introduces an error of order $\mathcal{O}(\delta t^2)$ per Floquet step relative to the Hamiltonian evolution, while $U_{\rm C}$ may also be regarded as a discrete-time circuit model in its own right.

\subsection{Circuit decompositions and resource estimates}
\label{Appendix:circuit decompositions subsection}

We now give the explicit gate decompositions used to implement the circuit families defined in Sec.~\ref{subsec: chiral circuit encoding}. The operator $u^{(0)}_n(\theta^{(u)}_n)$ can be implemented on IBM quantum hardware as the $XX+YY$ gate~\cite{google_sycamore},
\begin{equation}
\scalebox{0.9}{\begin{quantikz}
    &\gate[2]{XX+YY(\theta^{(u)}_n)}&\\
    &&\\
    \end{quantikz}} 
    = \begin{pmatrix}
        1 & 0 & 0 & 0 \\
        0 & \cos\frac{\theta^{(u)}_{n}}{2} & -i\sin\frac{\theta^{(u)}_{n}}{2} & 0 \\
        0 & -i\sin\frac{\theta^{(u)}_{n}}{2} & \cos\frac{\theta^{(u)}_{n}}{2} & 0 \\
        0 & 0 & 0 & 1 \\
    \end{pmatrix}
\end{equation}
whereas $u^{(2)}_n(\theta^{(v)}_n)$ is implemented by performing a $\pi/4$ $Z$ rotation on $u^{(0)}_n$ and conjugating it between two control-$Z$ and SWAP gates as
\begin{equation}
    \scalebox{0.79}{\begin{quantikz}
        & & &\gate[1]{R_z(-\pi/2)} & \gate[2]{XX+YY(\theta^{(v)}_{n})} &\gate[1]{R_z(\pi/2)} & & &\\
        & \ctrl{1} & \swap{1} & & && \swap{1} & \ctrl{1} &\\
        & \ctrl{-1} & \swap{-1} & & & & \swap{-1} & \ctrl{-1} &
    \end{quantikz}}.
\end{equation}
On IBM superconducting processors, two-qubit gates can only be applied between qubits connected in the hardware coupling map. The additional SWAP gates are therefore required to bring the relevant qubits into neighbouring positions before applying the controlled operation. On platforms with all-to-all connectivity, such as trapped-ion or neutral-atom processors, $u^{(2)}_n(\theta^{(v)}_n)$ could instead be implemented without these additional SWAP layers. For $N$ qubits and $M_{t}$ applications of $U_{\rm MF}$, the circuit can be implemented using $12NM_{t}$ controlled-$Z$ gates when using periodic boundary conditions, which is required for extracting the dispersion relation. On a platform with all-to-all connectivity, this count is reduced to $6NM_{t}$ controlled-$Z$ gates.

The interacting gates $u^{(1)}_n(\theta^{(v)}_n)$ and $u^{(3)}_n(\theta^{(v)}_n)$ admit related decompositions. Because of the location of the Pauli $Z$ operator in these terms, they can be implemented without the additional SWAP layer required for $u^{(2)}_n(\theta^{(v)}_n)$. The circuit decomposition of $u^{(1)}_n(\theta^{(v)}_n)$ is
\begin{equation}\label{eq:u1_gates}
    \scalebox{0.8}{\begin{quantikz}
        & \ctrl{1}& & & &\ctrl{1}  &\\
        & \ctrl{-1}  & \gate[1]{R_z(\pi/2)}& \gate[2]{XX+YY(
        \theta^{(v)}_{n})}&\gate[1]{R_z(-\pi/2)} & \ctrl{-1} &\\
        &  & & &  & &
    \end{quantikz}}
\end{equation}
while $u^{(3)}_n(\theta^{(v)}_n)$ can be decomposed as
\begin{equation} \label{eq:u3_gates}
    \scalebox{0.8}{\begin{quantikz}
        &  & \gate[1]{R_z(\pi/2)}& \gate[2]{XX+YY(\theta^{(v)}_{n})}&\gate[1]{R_z(-\pi/2)} &  &\\
        & \ctrl{1} & & &  & \ctrl{1}& \\
        & \ctrl{-1} & & &  & \ctrl{-1}& \\
    \end{quantikz}}
\end{equation}
These decompositions are adapted to the nearest-neighbour connectivity of the superconducting processor. For a circuit of $N$ qubits, $U_{\rm int}$ contains $6$ non-commuting Trotter layers per Floquet step. For $M_{t}$ applications with OBC, it can be implemented using $8(N-2)M_{t}$ controlled-$Z$ gates.

Finally, we consider the circuit complexity of the tunable version of $U_{\rm int}$ used in Sec.\ref{sec:scrambling}. Importantly, the circuit complexity depends on the chosen value of $\phi$. In the free-fermion limit, $\phi=0$, the circuit reduces to a modified XY-type circuit and can be implemented using $4NM_{t}$ controlled-$Z$ gates with PBC and $4(N-2)M_{t}$ with OBC. At $\phi=\pi$, the original interacting chiral circuit is recovered, requiring $8(N-2)M_{t}$ controlled-$Z$ gates. For generic values of $\phi$, each $CP(\phi)$ gate is decomposed using two controlled-$Z$ gates, giving a total complexity of $12(N-2)M_{t}$ with OBC. In Table.~\ref{table:comp table}, we give a full summary of the circuit complexities for unitaries considered throughout this work and for different boundary conditions.

\begin{table} 
\centering
\begin{tabular}{|c|c|c|}
\hline
Model & Boundary conditions &\#CZ gates\\
\hline
\hline
$U_{\text{MF}}$ & Open & $(12N-22)M_{t}$ \\
$U_{\text{MF}}$ & Periodic & $12NM_{t}$ \\
$U_{\text{XY}}$ & Open & $2(N-1)M_{t}$\\
$U_{\text{XY}}$ & Periodic & $2NM_{t}$\\
$U_{\phi=\pi}$ & Open, Periodic & $8(N-2)M_{t}$, $8NM_{t}$\\
$U_{\phi=0}$ & Open, Periodic & $4(N-2)M_{t}$, $4NM_{t}$\\
$U_{\phi}$ & Open, Periodic & $12(N-2)M_{t}$, $12NM_{t}$\\
\hline
\end{tabular}
\caption{Table demonstrating the circuit complexity of all models considered throughout this work. The model $U_{\phi}$, represents the scrambling unitary when $\phi\neq0,\pi$. The circuit's complexity is quantified in terms of controlled-$Z$ gates, which are the native two-qubit gate for \ibmboston{}. All complexities are given before optimisation by the \textit{Qiskit} transpiler.}
\label{table:comp table}
\end{table}

\subsection{Spectroscopic reconstruction of the dispersion}
\label{Appendix:spectroscopy}

To extract the single-particle dispersion relation, we prepare a coherent superposition of the vacuum and a single excitation. The qubit at site $N/2$ is initialised in the state $|+\rangle$, while all remaining qubits are prepared in $|0\rangle$, giving
\begin{equation}
|\psi(0)\rangle = \frac{1}{\sqrt{2}} \left( |00\cdots 000\cdots0\rangle + |00\cdots 010 \cdots0\rangle \right).
\end{equation}
The state is evolved under the mean-field circuit $U_{\rm MF}$ with periodic boundary conditions. The local coherence $\langle\sigma^+_j\rangle$ probes the amplitude and phase accumulated by the single excitation relative to the vacuum and is reconstructed from measurements in the $X$ and $Y$ bases using Eq.~\eqref{eq:sigmaplus}.

Expanding the evolved state in the momentum basis gives
\begin{equation}
    \langle\sigma^{+}_{j}\rangle = \frac{1}{2\sqrt{N}}\sum_{k}\alpha_{k}^{*}e^{i(\omega(k)t - kj)},
\end{equation}
where $\alpha_k$ is the momentum-space amplitude of the excitation. A two-dimensional Fourier transform with respect to position and time, therefore, resolves the dominant quasi-energies as a function of momentum. Although we focus here on the single-particle sector, the same protocol can be extended to few-particle spectroscopy by preparing and resolving states containing a small number of excitations.

\subsection{Hawking temperature circuits and decoherence normalisation}
\label{app:TH_decoherence}

\begin{figure}[ht!]
    \centering
\includegraphics[width=0.99\linewidth]{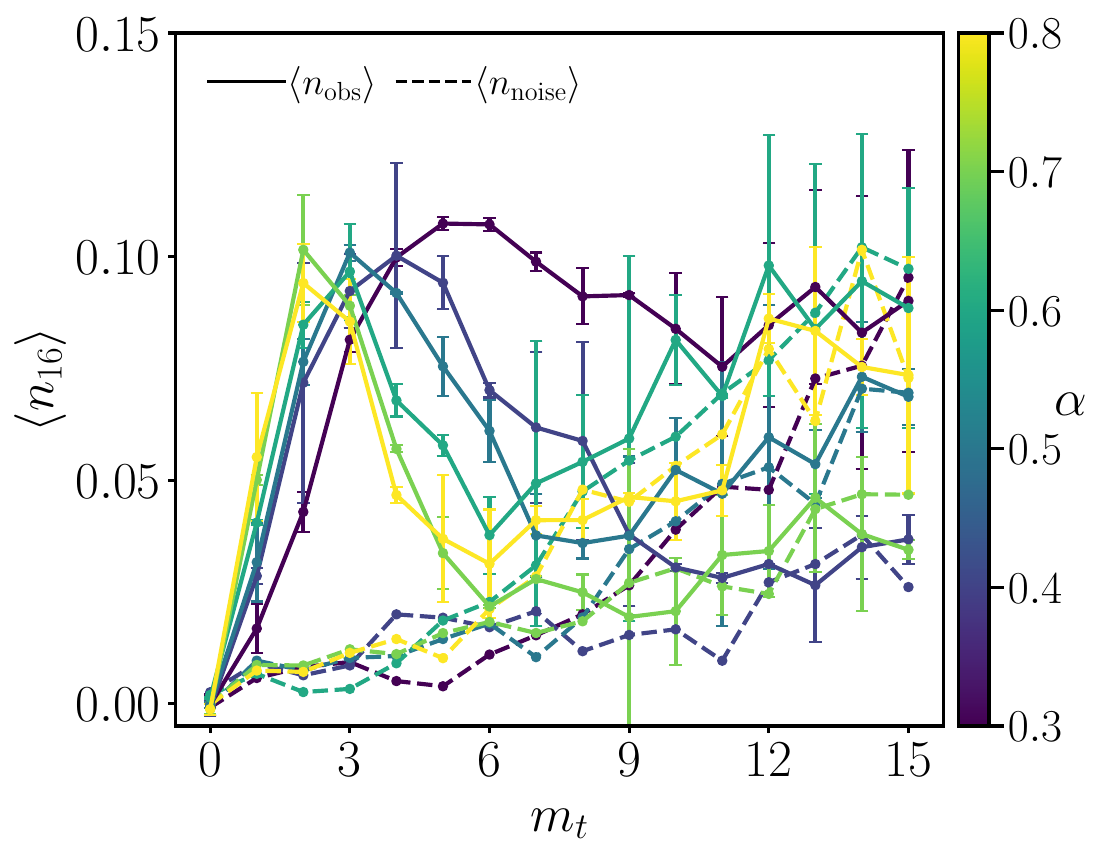}
    \caption{Demonstration of the error mitigation strategy for the Hawking temperature on \ibmboston{}. The solid and dashed lines show the hardware-measured population densities for the time-evolved excited and unexcited initial states, respectively, where the latter is used as a measure of the initial state's decoherence. Results are for $N=32$ qubits with the horizon positioned at $n_h=15.5$ and the initial single-particle excitation on site $j=15$. The population densities $\langle n_j\rangle$ are measured outside the horizon at site $j=16$.}
    \label{fig:hawking_decoherence}
\end{figure}

To quantify the impact of hardware-induced decoherence on the measured population density outside the horizon, we directly characterise the noise by applying the wave packet evolution protocol to the trivial initial state $\ket{00...0}$. Since this state contains no spin excitations, and the XY and chiral Hamiltonians are $U(1)$ conserving, the population density outside the horizon should be identically zero at all times. However, as shown in Fig.~\ref{fig:hawking_decoherence}, measurements on the quantum hardware exhibit a non-zero population density $\braket{n_\text{noise}}$ that increases over time, which arises due to the decoherence during circuit execution. We observe that the magnitude of this effect scales approximately linearly with the number of time cycles, reflecting increased qubit exposure to environmental noise. Consequently, later gate operations act on progressively decohered qubit states, leading to an accumulation of errors in the measured observables.

In the wave packet simulations, the initial excitations are prepared near the horizon, so that the measured population peaks occur after relatively short times and the influence of decoherence remains small. This ensures that the extracted Hawking temperatures are only minimally affected by hardware noise. To further reduce erroneous noise effects in our results, we normalise the measured population density $\braket{n_\text{obs}}$ of the time-evolved wave packet with that of the decohered unexcited state, $\braket{n_\text{noise}}$, as
\begin{equation}
    \braket{n_\text{norm}} = \bigg|\frac{\braket{n_\text{obs}}-\braket{n_\text{noise}}}{1-\braket{n_\text{noise}}}\bigg|.
    \label{Equation: Normalised Site Occupation Expectation Value}
\end{equation}
This normalisation mitigates the contribution of the wave packet's decoherence to the measured population density, allowing clearer resolution of the peaks at late times.

As the hardware measurement of $\braket{n_\text{obs}}$ and $\braket{n_\text{noise}}$ both incur errors, we quantify the error in the normalised population density using the standard deviation \cite{Taylor_1997}
\begin{align}
    \sigma_\text{norm} & = \sqrt{\left(\frac{\partial\braket{n_\text{norm}}}{\partial\braket{n_\text{obs}}}~\sigma_\text{obs} \right)^2 + \left(\frac{\partial\braket{n_\text{norm}}}{\partial\braket{n_\text{noise}}}~\sigma_\text{noise}\right)^2}, \\
    & = \frac{1}{1-\braket{n_\text{noise}}}\sqrt{\sigma_\text{obs}^2 + \sigma_\text{noise}^2 \left(\frac{\braket{n_{\text{obs}}}-1}{1-\braket{n_\text{noise}}}\right)^2} .
    \label{Equation:Variance of Normalised Site Occupation Expectation Value}
\end{align}

\subsection{OTOC echo circuit and reference normalisation}
\label{app:OTOC_protocol}

Here, we provide the circuit used to estimate the infinite-temperature OTOC
\begin{equation}
    C(t)=\frac{1}{2^N}\mathrm{Tr}\!\left[B(t)MB(t)M\right],
    \qquad
    B(t)=U_{\phi}^{\dagger}(t)BU_{\phi}(t).
    \label{eq:appendix_OTOC}
\end{equation}
Let $|m_+\rangle_i$ denote the $+1$ eigenstate of the probe operator $M_i$. For each random bitstring
\begin{equation}
    \mathbf{b} = \{b_r\}_{r\neq i}, 
    \qquad
    b_r\in\{0,1\},
\end{equation}
we prepare the product state
\begin{equation}
    |\psi_{\mathbf b}\rangle
    =
    |m_+\rangle_i
    \bigotimes_{r\neq i}|b_r\rangle .
    \label{eq:random_OTOC_state}
\end{equation}

The corresponding circuit is
\begin{equation}
\label{eq:OTOC_circuit_appendix}
\begin{quantikz} 
\lstick{$\ket{m_{+}}$} & \gate[4]{U_{\phi}(t)} & & \gate[4]{U_{\phi}(t)^\dagger} & \gate[1]{M}&\meter{} \qw \\ 
\lstick{$\ket{b_r}$} & & & & & \qw \\ 
\lstick{$\ket{b_r}$} & & \gate[1]{B} & & & \qw \\ 
\lstick{$\ket{b_r}$} & & & & & \qw 
\end{quantikz} .
\end{equation}

For a given bitstring, the measured echo signal is
\begin{equation}
    c_{\mathbf b}(t)
    =
    \langle\psi_{\mathbf b}|
    B(t)M_iB(t)
    |\psi_{\mathbf b}\rangle.
    \label{eq:single_bitstring_OTOC}
\end{equation}
As $M_i|\psi_{\mathbf b}\rangle=|\psi_{\mathbf b}\rangle$, this can equivalently be written as
\begin{equation}
    c_{\mathbf b}(t)
    =
    \langle\psi_{\mathbf b}|
    B(t)M_iB(t)M_i
    |\psi_{\mathbf b}\rangle.
\end{equation}

A uniform average over all computational-basis states of the remaining $N-1$ qubits gives
\begin{equation}
    \frac{1}{2^{N-1}}
    \sum_{\mathbf b}
    |\psi_{\mathbf b}\rangle
    \langle\psi_{\mathbf b}|
    =
    \frac{\mathds{1}+M_i}{2^N}.
\end{equation}
Since $B(t)^2=\mathds{1}$ and $\mathrm{Tr}(M_i)=0$, the term proportional to the identity does not contribute, yielding
\begin{equation}
    \frac{1}{2^{N-1}}
    \sum_{\mathbf b}
    c_{\mathbf b}(t)
    =
    \frac{1}{2^N}
    \mathrm{Tr}\!\left[
    B(t)M_iB(t)M_i
    \right]
    =
    C(t).
\end{equation}
The complete average is approximated experimentally using $N_r$ randomly sampled bitstrings,
\begin{equation}
    C_{\mathrm{raw}}(t)
    =
    \frac{1}{N_r}
    \sum_{\mathbf b}
    c_{\mathbf b}(t).
    \label{eq:sampled_OTOC}
\end{equation}

To estimate the attenuation produced by the hardware, we also execute the reference circuit with the butterfly operator omitted. In the absence of errors,
\begin{equation}
    U_{\phi}^{\dagger}(t)U_{\phi}(t)=\mathds{1},
\end{equation}
and the reference expectation value is therefore $C_{\mathrm{ref}}(t)=1$ for every evolution time.

On hardware, deviations of $C_{\mathrm{ref}}(t)$ from unity primarily reflect decoherence, imperfect gates and inaccuracies in reversing the evolution. We define the corrected correlator by
\begin{equation}
    C_{\mathrm{mit}}(t)
    =
    \frac{
        C_{\mathrm{raw}}(t)/C_{\mathrm{ref}}(t)
    }{
        C_{\mathrm{raw}}(0)/C_{\mathrm{ref}}(0)
    },
    \label{eq:OTOC_reference_normalisation}
\end{equation}
which ensures that $C_{\mathrm{mit}}(0)=1$. This correction assumes that the OTOC and reference circuits experience approximately the same background attenuation, an assumption justified by their identical many-body evolution and their difference of only a single local Pauli gate.

\section{Technical details of hardware simulations}\label{Appendix:technical details}

Throughout this work, we have demonstrated that superconducting quantum hardware is an excellent platform for studying the dynamical physics of black holes. Through a circuit decomposition of the spin-$1/2$ chiral spin-chain, we are able to experimentally probe both the Hawking temperature and scrambling physics. In this appendix, we provide the technical details of our experimental simulations implemented on the quantum hardware.

All the hardware simulations were performed on \ibmboston{} during the months of May to August. All the circuits were executed with $n_s=10000$ shots and performed using the \textit{EstimatorV2} primitive. The number of qubits and Trotter steps for each simulation can be found in their respective sections in the main text. Finally, we use three different error mitigation techniques to improve the outputted expectation values, which are Pauli twirling \cite{Wallman2016paulitwirling}, measurement readout twirling \cite{Karalekas2020measurementtwirling, Berg2022measurementtwirling} and zero noise extrapolations (ZNE) \cite{Temme2017ZNE}. For the Pauli twirling and measurement readout twirling methods, we allow \textit{Qiskit} to optimise for the total number of circuit randomisations and the allocated number of shots for each randomisation. For ZNE, we choose the noise amplification strategy of gate folding ~\cite{Giurgica2020Gatefolding} with noise factors of 1, 3 and 5 and extrapolated to the zero noise limit using the \textit{Qiskit Runtime's} exponential and linear extrapolation models. 

Finally, we consider the technical details for each of the three sets of hardware simulations: dispersion relation, Hawking temperature, and scrambling. Each set of hardware simulations contains a different parallelisation protocol and a two-qubit gate complexity, which we give in turn. 

\begin{figure}
    \centering
    \includegraphics[width=0.99\linewidth]{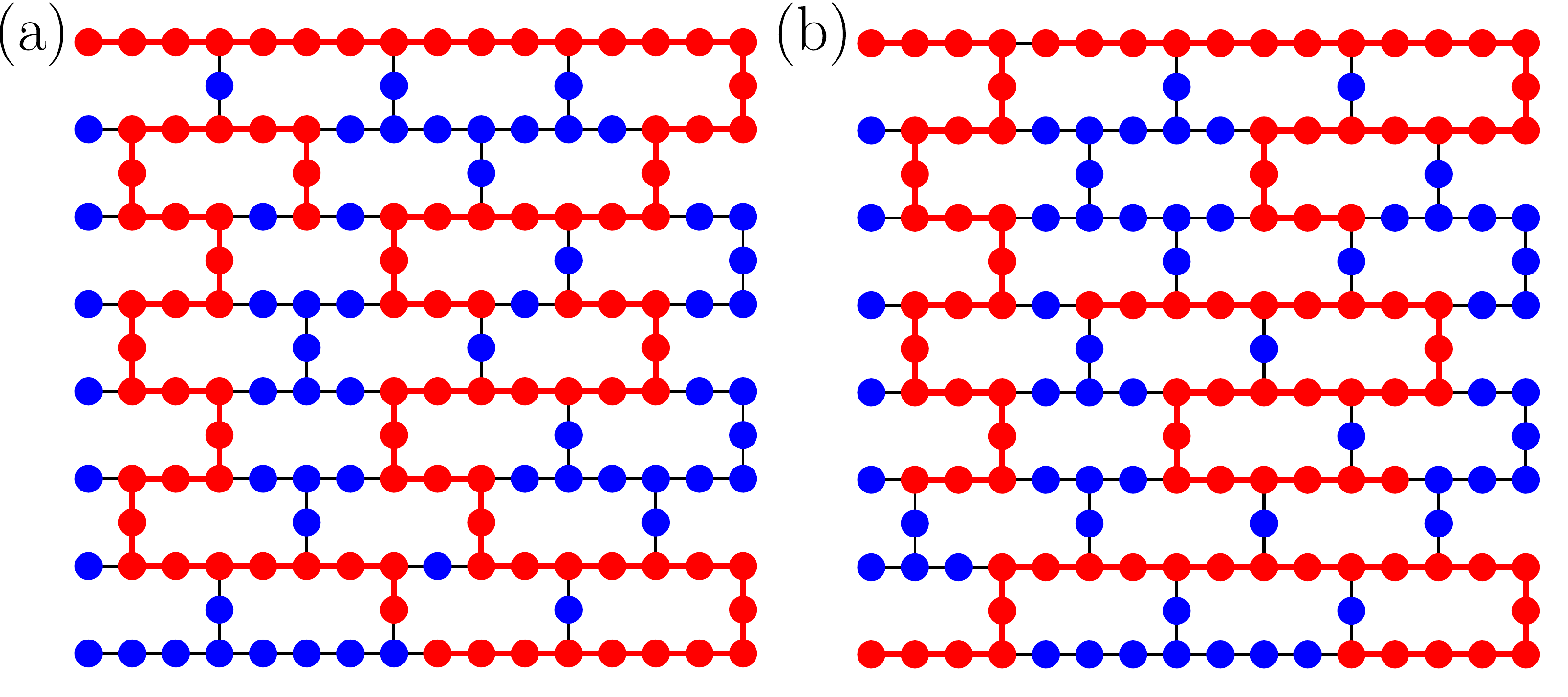}
    \caption{Chosen qubits for the different experiments on the superconducting processor. (a): Qubit mapping onto \ibmboston{} for the Hawking temperature experiments, which contain $3$ independent chains of $N=32$ qubits that are run in parallel, denoted by the red circles and connections. The blue circles and black lines are the unused qubits and connections. This mapping was used to produce Fig.~\ref{fig:XY_hawking}(b-c). (b): A similar qubit mapping, but for the scrambling circuits, which contain 4 chains of $N=24$ qubits and were used to produce Figures.~\ref{fig:hardware scrambling}(a-b).}
    \label{fig:Hardware layout}
\end{figure}

For the dispersion relation, we do not perform any parallelisation for the three different values of $\theta^{(v)}$ and obtain the experimental observables $\langle X_{i}\rangle$ and $\langle Y_{i}\rangle$ using a ring of $N=24$ qubits. For the largest circuit given at $M_t=20$ Trotter steps, the circuit implemented on the quantum processor contains $4099$ controlled-$Z$ gates after transpilation when $\theta^{(v)}\neq 0$. For the Hawking temperature, we parallelise the qubits into $3$ chains of $N=32$ qubits, allowing us to run all $6$ values of $\alpha$ using $2$ simulations. A diagram of this parallelisation on \ibmboston{} is presented in Fig.~\ref{fig:Hardware layout}(a), where the red circles and lines represent the qubits and connections used in the simulations. The blue circles and black lines represent unused qubits and connections. For a single value of $\alpha$, the largest circuit which contains $m_t=15$ Trotter steps has a circuit complexity of $930$ controlled-$Z$ gates after transpilation.

For the scrambling circuits, we can simulate 4 independent chains of $N=24$ qubits up to $M_{t}=10$ Trotter steps, allowing us to run all $20$ initial bitstrings over 5 runs. For the normalisation procedure given in Sec.~\ref {app:OTOC_protocol}, the same qubits are used for each corresponding bitstring. A similar qubit mapping onto \ibmboston{} is shown in Fig.~\ref{fig:Hardware layout}(b). While not given here, we also perform a similar qubit parallelisation for the $N=18$ scrambling circuits over $4$ independent chains. For the post-transpiled controlled-$Z$ circuit complexity, we need to consider two cases: the first is the complexity of the full echo protocol with the addition of the butterfly operator $B_{i}$, and the second is the circuits without $B_{i}$, which is used for mitigation. We note that for mitigated circuits, since $U_{\phi}^{\dagger}(t)U_{\phi}(t)=\mathds{1}$, we need to include a barrier operation to prevent the \textit{Qiskit} compiler from removing all gates during the transpilation process, which will lead to a higher circuit complexity. In the free fermion limit, $\phi=0$, the echo protocol will have $1262$ controlled-$Z$ gates while the largest mitigated circuit will have $1480$ controlled-$Z$ gates. At the chiral point, $\phi=\pi$, the standard and mitigated echo protocol will contain $3056$ and $3268$ controlled-$Z$ gates, respectively. Lastly, for all other values of $\phi$, the two echo protocols will have $4710$ and $5028$ controlled-$Z$ gates, respectively. All the pre-transpiled circuit controlled-$Z$ scaling for all the considered experiments can be found in Table.~\ref{table:comp table}. 

\section{Exact diagonalisation simulations}\label{sec:appendix_ED}

\subsection{Simulating Hawking temperature using Trotterised dynamics}

\begin{figure}[ht!]
    \centering
\includegraphics[width=0.99\linewidth]{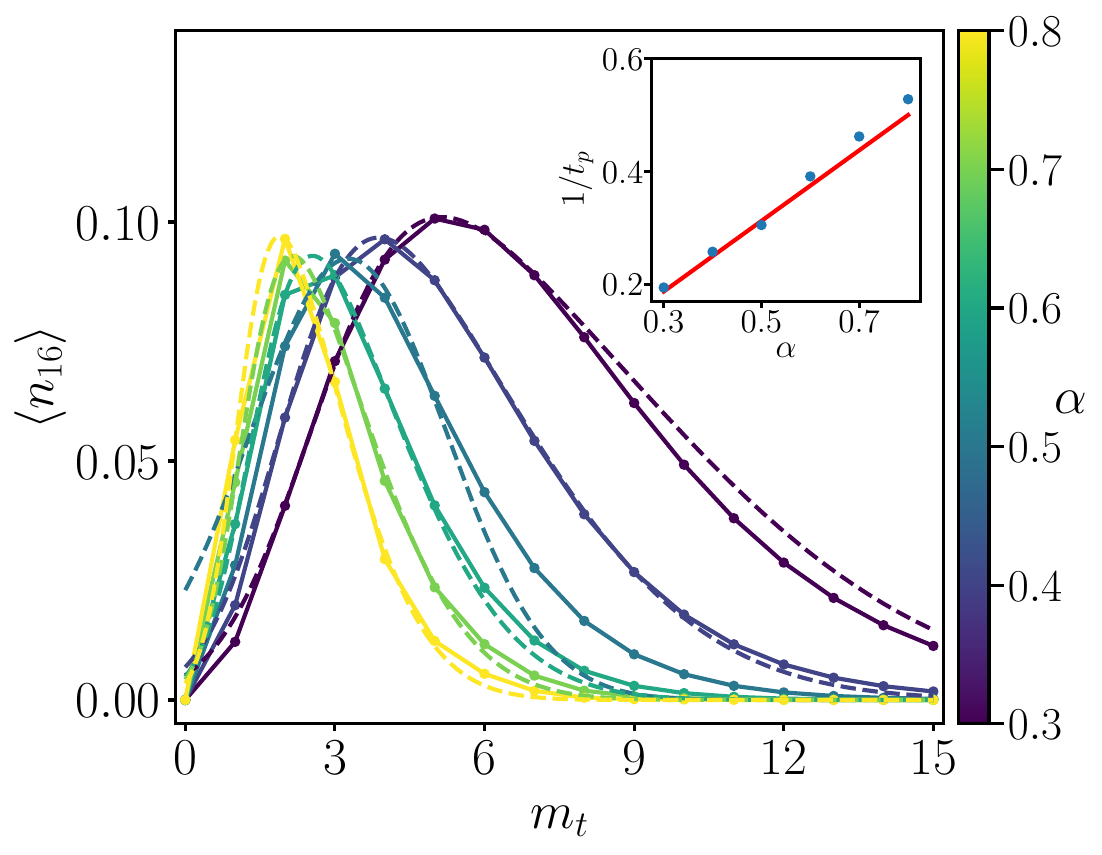}
    \caption{Exact diagonalisation results for the population density $\langle n_j \rangle$ following the time-evolution of the single-particle excitation under the XY Floquet unitary $U_{\rm XY}$. The excitation, initially localised at site $j=15$, tunnels across the horizon at $n_h=15.5$ to the exterior, resulting in a population density pulse computed at site $j=16$. Skew-Gaussians are fitted to the pulses to determine their characteristic peak time $t_p$. Inset: Numerical verification of the inverse relationship $t_p=8\gamma/ \alpha$, with $\gamma\approx0.186$, determined from the peak times.}
    \label{fig:hawking_ed}
\end{figure}
Here, the inverse scaling between the density pulse's characteristic peak time $t_p$ and the coupling profile's gradient at the horizon $\alpha$ is confirmed for the Trotterised dynamics using exact diagonalisation. Analogous to the method described in Sec.~\ref{sec:Hawking temperature}, an initial state $|\psi(0)\rangle$ of a single-particle excitation localised just behind the horizon $n_h=(N-1)/2$ is time-evolved using the XY Floquet unitary of Eq.~\eqref{eq:XYfloquetcircuit} to obtain $\ket{\psi(t)}$. At each Trotter step $m_t=t/\delta t$, the population density $\langle n_j (t) \rangle$ is computed for the exterior site adjacent to the horizon, $j=N/2$. Fig.~\ref{fig:hawking_ed} shows the numerically simulated population density for $N=32$ lattice sites with the coupling $u_{\rm XY}(x)=f(x)$ and metric profile of Eq.~\eqref{eqn:profile}. Fitting a skew-Gaussian to the observed pulses in the population density allows their peak times to be extracted, which are confirmed to scale as $t_p=8\gamma/\alpha$ with $\gamma= 0.186$, as illustrated in the inset. That the value of $\gamma$ obtained with the hardware is similar to that from exact diagonalisation indicates that, despite accumulating decoherence at late times, our error mitigation protocol allows the calibration constant $\gamma$ to be faithfully extracted.

\begin{figure}[ht!]
    \centering
\includegraphics[width=0.99\linewidth]{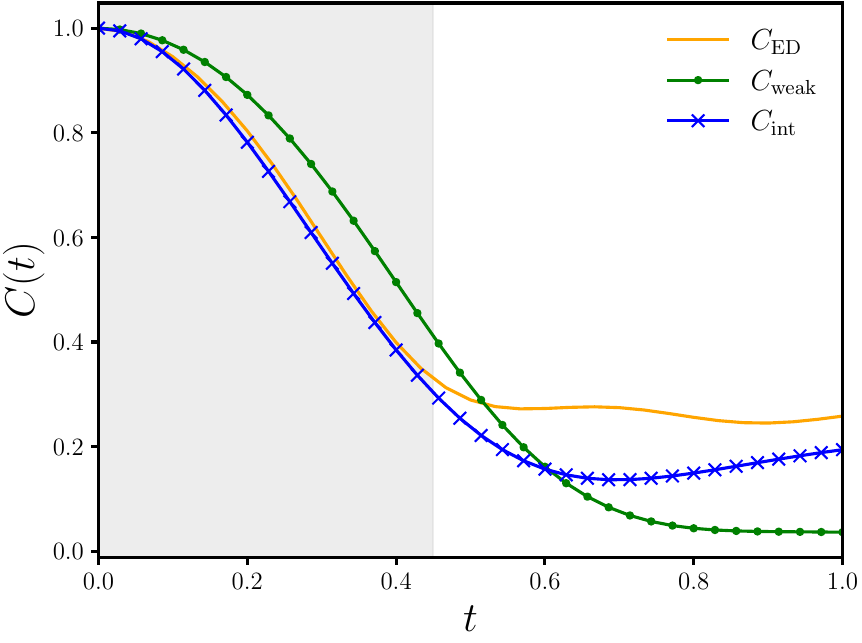}
    \caption{Comparison of the infinite temperature OTOCs for the dynamics generated by exact diagonalisation of the chiral Hamiltonian and the Trotterised gates $U_{\rm weak}$ and $U_{\rm int}$, for $N=12$ lattice sites with homogeneous couplings $u=0$ and $v=20$, and $M=35$ Trotter steps. Initial exponential chaotic decay of the chiral Hamiltonian's dynamics is replicated by the interacting dynamics of $U_{\rm int}$, as shown by the correlation between decays within the initial grey region. The dynamics of gates $u^{(2)}_n(\theta^{(v)}_n)$, which capture the mean-field Hamiltonian dynamics, show weaker decay, as expected from a system with fewer interactions.}
    \label{fig:gate_comparisons}
\end{figure}

\subsection{Trotterised dynamics as approximations of chiral scrambling}

As the Trotterisation of the Hamiltonians' dynamics in this work is an approximation, it is of interest to investigate how reliably this method replicates the chiral spin-chain's interactions. To determine the reproducibility of maximally chaotic dynamics, we compare exact diagonalisation results for the original infinite-temperature OTOCs, produced by the full chiral Hamiltonian of Eq.~\eqref{eq:chiralHamiltonian}, with those obtained using the Trotterised gate sequence, and extract the Lyapunov exponents to evaluate the difference in scrambling.

Fig.~\ref{fig:gate_comparisons} shows the OTOCs computed for a lattice of $N=12$ sites for time-evolution generated by the Hamiltonian dynamics and Trotterised gate sets 
\begin{equation}
\begin{split}
& U_{\rm weak} =\\
&
\prod_{\scriptscriptstyle n\equiv2\,(\mathrm{mod}\,3)} u^{(2)}_n(\theta^{(v)}_n)
\prod_{\scriptscriptstyle n\equiv1\,(\mathrm{mod}\,3)} u^{(2)}_n(\theta^{(v)}_n)
\prod_{\scriptscriptstyle n\equiv0\,(\mathrm{mod}\,3)} u^{(2)}_n(\theta^{(v)}_n)
\\
&\times
\prod_{\scriptscriptstyle n\equiv2\,(\mathrm{mod}\,3)} u^{(1)}_n(\theta^{(v)}_n)
\prod_{\scriptscriptstyle n\equiv1\,(\mathrm{mod}\,3)} u^{(1)}_n(\theta^{(v)}_n)
\prod_{\scriptscriptstyle n\equiv0\,(\mathrm{mod}\,3)} u^{(1)}_n(\theta^{(v)}_n). \label{eq:weak_int_gate_structure}
\end{split} 
\end{equation}
and $U_{\rm int}$ of Eq.~\eqref{eq:int_gate_structure}. Whilst the $U_{\rm int}$ gate set produces the interacting contributions to the chiral spin-chain's dynamics, the gates $u^{(2)}_n(\theta^{(v)}_n)$ in $U_{\rm weak}$ are responsible for the mean-field dynamics of the chiral operator. Thus, it is expected that the chaotic features of the dynamics generated by $U_{\rm weak}$ will be noticeably weaker. As expected, Fig.~\ref{fig:gate_comparisons} demonstrates that the initial exponential decay of the chiral Hamiltonian's dynamics (shown as the shaded portion of the figure) is reliably reproduced from the Trotterised dynamics of $U_{\rm int}$, with fittings of the hypergeometric function Eq.~\eqref{eq:kummer} giving fitted exponents of $\lambda_{\rm ED}/v=0.78$ and $\lambda_{\rm int}/v=0.80$, respectively. In contrast, the dynamics produced by $U_{\rm weak}$ display a smaller initial decay with $\lambda_{\rm weak}/v=0.70$, indicating slower information scrambling and weaker chaotic potential than the interacting dynamics.

\bibliography{references}

\end{document}